\documentclass[amsmath,trackchanges,twocolumn]{aastex701}
\usepackage{float}

\usepackage{booktabs}
\usepackage{amsmath}
\usepackage{makecell}
\usepackage{subfigure}
\usepackage{multirow}
\usepackage{threeparttable}
\usepackage{hyperref}

\begin{document}

\title{Locating the Production Sites of High-Energy Neutrinos in Blazar Jets}

\author[orcid=0000-0003-1721-151X,gname=Rui,sname=Xue]{Rui Xue}
\affiliation{Department of Physics, Zhejiang Normal University, Jinhua 321004, China}
\email[show]{ruixue@zjnu.edu.cn}

\author[0000-0002-7272-1136,gname=Yoshiyuki,sname=Inoue]{Yoshiyuki Inoue}
\affiliation{College of Systems Engineering and Science, Shibaura Institute of Technology, 307 Fukasaku, Minuma-ku, Saitama City, Saitama 337-8570, Japan}
\affiliation{Interdisciplinary Theoretical \& Mathematical Science Center (iTHEMS), RIKEN, 2-1, Hirosawa, Wako, Saitama 351-0198, Japan}
\affiliation{Kavli Institute for the Physics and Mathematics of the Universe (WPI), The University of Tokyo, Kashiwa 277-8583, Japan}
\email[show]{yinoue@shibaura-it.ac.jp}

\author[orcid=0000-0002-3883-6669,gname=Ze-Rui,sname=Wang]{Ze-Rui Wang} 
\affiliation{College of Physics and Electronic Engineering, Qilu Normal University, Jinan 250200, China}
\affiliation{Shandong Key Laboratory of Space Environment and Exploration Technology, China}
\email{zerui\_wang62@163.com}

\author[orcid=0000-0001-6614-3344,gname=Neng-Hui,sname=Liao]{Neng-Hui Liao} 
\affiliation{Department of Physics and Astronomy, College of Physics, Guizhou University, Guiyang 550025, People's Republic of China}
\email{nhliao@gzu.edu.cn}

\author[0000-0002-6809-9575,gname=Dingrong,sname=Xiong]{Dingrong Xiong}
\affiliation{Yunnan Observatories, Chinese Academy of Sciences, 396 Yangfangwang, Guandu District, Kunming, 650216, China}
\affiliation{Center for Astronomical Mega-Science, Chinese Academy of Sciences, 20A Datun Road, Chaoyang District, Beijing, 100012, China}
\affiliation{Key Laboratory for the Structure and Evolution of Celestial Objects, Chinese Academy of Sciences, 396 Yangfangwang, Guandu District, Kunming, 650216, China}
\email[show]{\\xiongdingrong@ynao.ac.cn}

\begin{abstract}
The production sites of high-energy neutrinos in blazar jets remain poorly constrained. In this work, we investigate the physical conditions required for efficient neutrino production by using radio-constrained jet properties to evaluate the radial evolution of the external-to-magnetic energy density ratio (Compton dominance $Y$). We identify $Y \gg 1$ as the key physical condition for efficient neutrino production, as it simultaneously enhances photohadronic interactions and suppresses synchrotron radiation from secondary pairs, thereby avoiding an excess of hard X-ray emission. We find that such large values of $Y$ are most readily achieved near or within the broad-line region. This large-$Y$ condition is generally incompatible with reproducing the observed broadband spectral energy distribution within a single emission region, naturally indicating that the neutrino-emitting region is physically distinct from the dominant electromagnetic emission zone.
We further show that such a scenario can be realized either if the jet completes its acceleration within sub-parsec scales or if the bulk Lorentz factor is intrinsically large, both of which appear uncommon based on current observations. These results offer a physically motivated framework for identifying neutrino production sites, provide a natural explanation for the rarity of blazar--neutrino associations, and underscore the importance of constraining jet property at sub-parsec scales in the search for neutrino-emitting blazars.
\end{abstract}



\section{Introduction}  
The detection of the first astrophysical PeV-scale neutrinos by IceCube Neutrino Observatory in 2013 marked the advent of high-energy neutrino astronomy and triggered extensive efforts to identify their origins \citep{2013Sci...342E...1I, 2013PhRvL.111b1103A}. The nearly isotropic distribution of these neutrinos across the sky suggests an extragalactic origin \citep{2014PhRvL.113j1101A, 2017ApJ...849...67A, 2024NatAs...8..241F}. Among the various candidates, blazar jets have emerged as a promising candidate. This perspective was strengthened on 22 September 2017, when IceCube reported the first association between a high-energy neutrino IC-170922A and an extragalactic source, namely the blazar TXS\,0506+056, with a significance $\sim 3\sigma$ \citep{2018Sci...361.1378I}. Since then, a number of blazars have been identified as potential neutrino point sources, though only a few associations have reached significances of $\gtrsim3\sigma$ \citep[e.g.,][]{2018Sci...361..147I, 2020PhRvL.124e1103A}. Furthermore, the currently identified neutrino candidate blazars represent only a minor fraction of the overall blazar population. It also has been suggested that the spectral energy distributions (SEDs) and broadband properties of neutrino candidate blazars do not exhibit significant differences compared to those of other blazars \citep{2022MNRAS.510.2671P, 2023MNRAS.521.2270P, 2023MNRAS.526..661K}. This raises a fundamental question of what physical properties distinguish neutrino-emitting blazar jets from the majority of apparently neutrino-quiet blazars. 

Population studies have sought to identify which electromagnetic bands may serve as diagnostics or predictors of bright high-energy neutrino emission. The production of ultra-high-energy secondary particles in hadronic processes naturally links $\gamma$-ray emission to neutrino radiation. However, no clear correlation has been established for the $\gamma$-ray bright blazars so far \citep{2017ApJ...835...45A, 2019ICRC...36..916H, 2022ApJ...938...38A}. In contrast, possible correlations have been suggested in the radio (\citealt{2020ApJ...894..101P}; \citealt{2021ApJ...908..157P}; \citealt{2021A&A...650A..83H}; \citealt{2023MNRAS.523.1799P}, cf.,\ \citealt{Zhou:2021rhl, IceCube:2023htm, 2026ApJ...999...98A}) and hard X-ray \citep{2024JCAP...05..133P} bands. Recent work further suggests that blazars exhibiting extreme jet beaming, as inferred from radio observations, are statistically more likely to be associated with high-energy neutrinos \citep{2025ApJ...991...33P}. Population studies suggest that blazars as a class cannot fully account for the diffuse neutrino background \citep{2017ApJ...835...45A, 2019ICRC...36..916H, 2022ApJ...938...38A, 2024ApJ...973...97A}. This implies that only a minority of blazars are efficient neutrino emitters, suggesting substantial diversity in neutrino production ability among different blazar jets. Understanding whether such diversity originates from differences in intrinsic jet physical properties requires confronting theoretical models with observational constraints.

On the theoretical side, the one-zone jet model, widely adopted in blazar studies, has been extensively used to reproduce multiwavelength (MWL) electromagnetic emission and to account for the observed neutrino flux \citep[e.g.,][]{2014PhRvD..90b3007M, 2014JHEAp...3...29D, 2018ApJ...864...84K,2019MNRAS.483L..12C,2019NatAs...3...88G,2026ApJ..1005...30X}. However, constraints from hard X-ray data strongly suppress the synchrotron radiation of secondary pairs, whose luminosity would be comparable to the neutrino emission \citep[e.g.,][]{2018ApJ...865..124M}. Consequently, even under optimistic results, one-zone models face severe challenges in reproducing the observed neutrino flux \citep{2026ApJ..1005...45L}. In other words, the one-zone framework struggles to convincingly explain the associations, unless they are all attributed to statistical effects such as the Eddington bias \citep{2019A&A...622L...9S}. Therefore, hard X-ray constraints indicate that the key challenge is not simply reproducing the observed neutrino flux, but identifying the physical conditions under which secondary synchrotron emission can be efficiently suppressed while maintaining efficient neutrino production \citep{2019ApJ...886...23X}.

Building on this insight, multi-zone approaches, such as the inner--outer blob jet model \citep{2019PhRvD..99f3008L,2019ApJ...886...23X,2021RAA....21..305W,2021ApJ...906...51X}, have been proposed to enhance the model-predicted neutrino flux while remaining consistent with hard X-ray constraints. A key feature of these models is the assumption of a compact neutrino-emitting region with a large Compton dominance, $Y = U'_{\rm ext}/U'_{\rm B} \gg 1$, where $U'_{\rm ext}$ and $U'_{\rm B}$ denote the external photon and magnetic energy densities in the jet comoving frame. Under such conditions, secondary pairs (produced by $\pi^{\pm}$ decay, Bethe--Heitler pair production, and $\gamma\gamma$ annihilation) mainly cool via external Compton (EC) scattering rather than synchrotron radiation, thereby alleviating the hard X-ray constraints while simultaneously enhancing photohadronic interactions \citep{2019ApJ...886...23X}. However, the drawback of multi-zone models is the introduction of additional free parameters, and the required $Y \gg 1$ condition is often realized through fine-tuning. Consequently, whether and where such a large-$Y$ environment can be realized in realistic blazar jets remains a central question in establishing a robust physical explanation of the neutrino--blazar associations.

In this work, we address this question by employing radio-constrained jet parameters to locate regions capable of achieving $Y \gg 1$, thereby identifying potential sites of high-energy neutrino production. Although radio observations primarily trace parsec scale regions, they provide robust constraints on global jet beaming and magnetic field strength, which can be extrapolated to constrain the properties of the inner dissipation regions. Since the $Y$ parameter depends on the ratio of external photons to magnetic energy densities, independent radio constraints on the jet beaming factor and magnetic field strength directly test whether the extreme condition $Y \gg 1$ can be physically realized. Investigating where this condition can be realized under realistic jet parameters naturally identifies the jet regions capable of efficient high-energy neutrino production, thereby providing a physically motivated framework for locating the neutrino production sites. This paper is organized as follows. In Sect.\,\ref{sec:method}, we establish a theoretical framework to evaluate the external photon fields and jet physical parameters using MWL and radio observations. This framework is subsequently applied to TXS\,0506+056 and PKS\,1502+106 in Sect.\,\ref{sec:app} to identify potential neutrino production sites and perform self-consistent SED modeling. We then extend the analysis to a broader blazar population based on the MOJAVE sample in Sect.\,\ref{sec:popu}, where we construct a proxy for $Y$ and identify promising neutrino-emitting candidates. Our conclusions are presented in Sect.\,\ref{sec:con}. Throughout the paper, the cosmological parameters $H_{0}=69.6\ \rm km\ s^{-1}Mpc^{-1}$, $\Omega_{0}=0.29$, and $\Omega_{\Lambda}$= 0.71 are adopted \citep{2014ApJ...794..135B}.

\section{Method} \label{sec:method}
There is growing evidence that low- and intermediate-synchrotron-peaked (LSP/ISP) blazars \citep{2010ApJ...716...30A}, as well as masquerading BL Lac objects \citep[mas BL Lacs;][]{2013MNRAS.431.1914G}, are more likely to be connected with high-energy neutrinos \citep{2019MNRAS.484.2067R, 2020ApJ...894..101P, 2021A&A...650A..83H, 2021ApJ...908..157P, 2022MNRAS.510.2671P, 2023MNRAS.521.2270P, 2023MNRAS.526..661K, 2023MNRAS.523.1799P, 2025A&A...700A.228A, 2025A&A...704A.184M, 2026APh...17503191V}. The key factor is that jets in LSP/ISP blazars and mas BL Lacs are generally believed to be embedded in abundant external photon fields, whose crucial role in neutrino production has been highlighted in many theoretical studies \citep{2019ApJ...886...23X, 2021ApJ...906...51X, 2024ApJ...962..142W, 2025PhRvD.112h3016W, 2024A&A...689A.147R, 2026A&A...706A.351R, 2026ApJ..1005...45L}, as their energy densities are boosted in the jet comoving frame, thereby enhancing the efficiency of the $p\gamma$ interaction. In this section, primed quantities are measured in the jet comoving frame, unless specified.
\subsection{Energy Densities of External Photons}\label{ext}
As suggested by the inner--outer blob model \citep[e.g.,][]{2019ApJ...886...23X}, the condition $Y \gg 1$ appears to be required in order to produce a neutrino flux sufficient to account for the observations without overshooting the hard X-ray constraints. A quantitative evaluation of the external radiation fields is essential. Here, we evaluate the external photon fields that may contribute to $U'_{\rm ext}$, including the accretion disk (AD), the X-ray hot corona (HC), the broad-line region (BLR), and the dusty torus (DT). 

Here we consider a standard disk model in which the AD is assumed to be geometrically
thin and optically thick \citep{1973A&A....24..337S}. Its emission profile has a multi-temperature blackbody shape, with the radial dependence of the temperature given by
\begin{equation}\label{eq14}
T(R\mathrm{_{AD}})=\left\{\frac{3R\mathrm{_{S}}L\mathrm{_{AD }}}{16\pi\eta\mathrm{_{acc}}\sigma\mathrm{_{SB}}R^{3}_{\rm AD}}\left [ 1-\left(\frac{3R\mathrm{_{S} }}{R\mathrm{_{AD}}}\right)^{1/2}\right ]\right\}^{1/4},
\end{equation}
where $R_{\rm S}$ is the Schwarzschild radius of the central supermassive black hole (SMBH), $L\mathrm{_{AD }}$ is the luminosity of the accretion disk, $R_{\rm AD}$ is the AD radius, ranging from $R_{\rm AD,in}=3R_{\rm S}$ to $R_{\rm AD,out}=500R_{\rm S}$, $\sigma\mathrm{_{SB}}$ is the Stefan-Boltzmann constant, $\eta\mathrm{_{acc}}$ is the accretion efficiency, adopted as 0.1. The energy density of the AD $U'_{\rm AD}$ in the jet comoving frame is \citep{2009MNRAS.397..985G}
\begin{equation}
U'_{\rm AD}(r_{\rm diss})=\frac{2\pi}{c}\int\int_{\mu_{\rm AD}}^1 \frac{I(\nu)}{\delta_{\rm AD}} d\mu d\nu'.
\end{equation}
Here, parameters are defined as follows. $\mu=r_{\rm diss}/\sqrt{r_{\rm diss}^2+R_{\rm AD}^2}$, where $r_{\rm diss}$ is the distance between the jet dissipation region and the SMBH; $\mu_{\rm AD}=(1+R_{\rm AD,out}^2/r_{\rm diss}^2)^{-1/2}$; $c$ is the speed of light; $\nu=\nu'\delta_{\rm AD}$ is the photon frequency in the SMBH frame; $I(\nu)=\frac{2h\nu^3/c^2}{e^{h\nu/kT}-1}$ is the AD emission intensity in the SMBH frame, where $k$ is the Boltzmann constant, and $h$ is the Planck constant; and $\delta_{\rm AD}=[\Gamma(1-\beta \mu)]^{-1}$, where $\Gamma$ is the jet bulk Lorentz factor and $\beta c$ is the corresponding velocity.

Surrounding the AD, there is a HC that emits X-ray radiation. The HC spectrum is described by a cut-off power-law of the form $L_\nu(\nu) \propto \nu^{-\alpha_{\rm HC}} \exp(-\nu/\nu_{\rm c})$, where $\alpha_{\rm HC}=1$ and $\nu_{\rm c}=150\,\mathrm{keV}/h$ are adopted \citep[e.g.,][]{2018MNRAS.480.1819R}. For simplicity, the HC is assumed to be spherical, with a radius $R_{\rm HC}=30R_{\rm S}$ \citep{2018MNRAS.480.1247K, 2018ApJ...869..114I}. The energy density of the HC $U'_{\rm HC}$ in the jet comoving frame can be evaluated as \citep{1996MNRAS.280...67G}
\begin{equation}
U'_{\rm HC}(r_{\rm diss})=\frac{f_{\rm X}L_{\rm AD}\Gamma^2}{4\pi R_{\rm HC}^2c}\big[1-\mu_{\rm X}-\beta(1-\mu_{\rm X}^2)+\frac{\beta^2}{3}(1-\mu_{\rm X}^3)\big],
\end{equation}
where $f_{\rm X}$ is assumed to be 0.1 \citep{2010A&A...512A..34L}, and $\mu_{\rm X}=(1+R_{\rm HC}^2/r_{\rm diss}^2)^{-1/2}$.

For the BLR and DT, they are assumed to be spherical shells with radii $R_{\rm BLR/DT}\gg R_{\rm AD}$. Following the reverberation mapping results \citep[e.g.,][]{2011A&A...536A..78K, 2013ApJ...767..149B}, their radii scale with the square root of $L_{\rm AD}$, i.e., $R_{\rm BLR/DT}=\xi_{\rm BLR/DT}(L_{\rm AD}/10^{46}\,\rm erg\,s^{-1})^{1/2}\,pc$, where $\xi_{\rm BLR}=0.1$ and $\xi_{\rm DT}=2.5$ \citep{2008MNRAS.387.1669G}. The energy densities of the BLR and DT $U'_{\rm BLR/DT}$ in the jet comoving frame can be evaluated as \citep{2009MNRAS.397..985G}
\begin{multline}
U'_{\rm BLR/DT}(r_{\rm diss})=
\frac{f_{\rm BLR/DT}L_{\rm AD}\Gamma^2}{4\pi R_{\rm BLR/DT}^2c}\\
\times 
\begin{cases}
1, & r_{\rm diss} \leqslant R_{\rm BLR/DT}, \\[6pt]
\left(\dfrac{r_{\rm diss}}{R_{\rm BLR/DT}}\right)^{-\alpha}, 
&\hspace{-55pt} R_{\rm BLR/DT} < r_{\rm diss} < 3\,R_{\rm BLR/DT}, \\[8pt]
\begin{aligned}
\dfrac{1}{3\beta}\big[&2(1-\beta \mu_1)^3-(1-\beta \mu_2)^3 \\
&-(1-\beta)^3 \big],
\end{aligned}
& r_{\rm diss} \geqslant 3\,R_{\rm BLR/DT},
\end{cases}
\end{multline}
where $f_{\rm BLR}$ and $f_{\rm DT}$ are taken to be 0.1 and 0.2, respectively \citep{2020NatCo..11.5475H}, $\mu_1=(1+R_{\rm BLR/DT}^2/r_{\rm diss}^2)^{-1/2}$, and $\mu_2=(1-R_{\rm BLR/DT}^2/r_{\rm diss}^2)^{1/2}$. In the intermediate region $R_{\rm BLR/DT} < r_{\rm diss} < 3R_{\rm BLR/DT}$, we use a power-law interpolation, with the index $\alpha$ chosen to ensure continuity at $r_{\rm diss}=3R_{\rm BLR/DT}$ \citep{2009MNRAS.397..985G}. The BLR and DT spectra are approximated as blackbody emission, peaking in the AGN frame at $2\times10^{15}\,\mathrm{Hz}$ and $3\times10^{13}\,\mathrm{Hz}$, respectively \citep{2008MNRAS.387.1669G}.

As described above, the external photon energy densities from the AD, HC, BLR, and DT can be calculated as functions of $r_{\rm diss}$. It can be seen that once $L_{\rm AD}$ is determined observationally, $\Gamma$ becomes the key parameter governing the transformation of these radiation fields into the jet comoving frame. As a result, $\Gamma$ determines the profiles of $U'_{\rm ext}(r_{\rm diss})$ along the jet.

\subsection{Radio Constraints on the Jet Beaming Factor and Magnetic Field}\label{radio}
To evaluate the ratio $Y=U'_{\rm ext}/U'_B$, constraints on both $\Gamma$ and the magnetic field strength $B'$ are required. In modeling studies, these two parameters are usually treated as free parameters. Radio observations, however, provide independent constraints on them from measurements of the parsec scale jet. In particular, measurements of the radio core brightness temperature constrain the jet beaming factor $\delta$, which in turn places limits on $\Gamma$\footnote{For blazars with a small viewing angle $\theta^{\rm obs} \lesssim 1/\Gamma$, we have $\delta \approx \Gamma$. Nevertheless, the framework is fully general. For misaligned AGN (e.g., radio galaxies) with larger viewing angles, $\Gamma$ can still be constrained, provided that high-resolution radio observations independently constrain both $\delta$ and apparent jet speed.}, while core-shift measurements provide estimates of $B'$ in the radio core region. With $\Gamma$ and $B'$ constrained, $Y$ can be evaluated along the jet to identify regions where the condition $Y\gg1$ could be satisfied.

In radio observations, the flux density is often parameterized in terms of the brightness temperature, $T_{\rm b}$. When the size of the emission region is determined through direct imaging, the observed radio core brightness temperature $T_{\rm b}^{\rm obs}$ is related to the intrinsic value $T'_{\rm b}$ as $T_{\rm b}^{\rm obs} = \delta_{\rm C}\, T'_{\rm b}$, with $\delta_{\rm C}$ being the Doppler factor at the radio core. The intrinsic brightness temperature is commonly assumed to be the equipartition value, $T_{\rm b,eq}=5\times10^{10}\,\rm K$, as defined by \cite{1994ApJ...426...51R}. Adopting $T_{\rm b,eq}$ as an upper limit for $T_{\rm b}$ then yields a lower limit on $\delta_{\rm C}$.
Radio observations show that AGN jets exhibit a parabolic geometry in the inner region near the jet base and transition into a conical structure at larger distances \citep{2012ApJ...745L..28A, 2020MNRAS.495.3576K}. This morphology is commonly interpreted as evidence that the inner jet is Poynting dominated, where the jet is gradually accelerated until the plasma approaches equipartition and reaches a terminal bulk Lorentz factor \citep{1977MNRAS.179..433B,2006MNRAS.368.1561M,2006MNRAS.367..375B,2013ApJ...775..118N}. Beyond the transition position, the jet becomes approximately conical and undergo gradual deceleration on larger scales, as suggested by observations \citep{2009MNRAS.398.1989M,2010ApJ...710..743D,2012ApJ...745L..28A,2015ApJ...798..134H}. Motivated by this picture, we adopt a simple parameterized jet acceleration profile, in which the jet first accelerates in the inner region and then decelerates slowly downstream. Therefore $\delta_{\rm C}$ provides a reference point for the jet dynamics, allowing $\delta(r_{\rm diss})$, and hence $\Gamma(r_{\rm diss})$, along the jet to be estimated. Here we assume \citep{2006MNRAS.367..375B, 2013MNRAS.429.1189P,2022PhRvD.105b3005W}
\begin{equation}\label{profile}
\Gamma(r_{\rm diss})=
\begin{cases}
\Gamma_{\rm tran}\left(\dfrac{r_{\rm diss}}{r_{\rm tran}}\right)^{1/2}, & r_{\rm diss}<r_{\rm tran}, \\[10pt]
\Gamma_\ast-\dfrac{\Gamma_\ast-2}{\log_{10}\!\left(\dfrac{r_{\max}}{r_\ast}\right)}
\log_{10}\!\left(\dfrac{r_{\rm diss}}{r_\ast}\right), & r_{\rm diss}\ge r_{\rm tran},
\end{cases}
\end{equation}
where $r_\ast = \max\{r_{\rm C},\,r_{\rm tran}\}$, $\Gamma_\ast = \Gamma_{\rm C}\left(\frac{r_\ast}{r_{\rm C}}\right)^{1/2}$, $\Gamma_{\rm tran}=\Gamma_\ast-\dfrac{\Gamma_\ast-2}{\log_{10}\!\left(\dfrac{r_{\max}}{r_\ast}\right)} \log_{10}\!\left(\dfrac{r_{\rm tran}}{r_\ast}\right)$, $\Gamma_{\rm C}=\delta_{\rm C}$ is the bulk Lorentz factor of the radio core located at $r_{\rm C}$, $r_{\rm tran}$ denotes the distance where the jet transitions from acceleration to deceleration, and $r_{\max}$ is the jet length, which we set to $10\,{\rm kpc}$ for simplicity.

Multi-frequency high-angular-resolution radio observations of the core-shift effect provide a method to estimate the radial evolution of $B'(r_{\rm diss})$ \citep{1998A&A...330...79L, 2005ApJ...619...73H}. Assuming equipartition between the magnetic field and non-thermal relativistic particle energy densities at $r_{\rm C}$, the magnetic field strength at 1 pc, $B'_{\rm 1\,pc}$, can be estimated through \citep{2009MNRAS.400...26O, 2014Natur.510..126Z}
\begin{equation}\label{b1pc}
\begin{split}
B'_{1\,\mathrm{pc}} \;&\approx\; 0.025 \left( \sigma_{\rm rel}
\frac{\Omega_{r\nu}^{3} (1+z)^{2}}{\delta^{2} \, \phi \, \sin^{2}\theta}
\right)^{1/4}\\
&\approx \sigma_{\rm rel}^{1/4} B'^{\rm B,e}_{\rm 1\,pc},
\end{split}
\end{equation}
where $z$ is the redshift, $\Omega_{r\nu}$ is the core-shift measure, $\theta$ is the viewing angle, and $\phi$ is the jet half-opening angle. Here $\sigma_{\rm rel}$ denotes the ratio of the magnetic-field energy density to that of the non-thermal relativistic particles, and $B'^{\rm B,e}_{\rm 1\,pc}$ represents the value typically derived from radio studies that generally adopt $\sigma_{\rm rel}=1$, i.e., equipartition reached between relativistic electrons and magnetic field. Throughout this work, we adopt $\sigma_{\rm rel}=10^{-4}$, i.e., equipartition reached between relativistic protons and magnetic field, a typical value for hadronic jet models \citep[e.g.,][]{2013ApJ...768...54B}.
Although the magnetic field configuration in the jet acceleration and dissipation region, particularly on sub-parsec scales, remains poorly constrained and is likely more complex than a simple self-similar structure, observations across multiple wavebands, probing jet regions from parsec to sub-parsec scales, generally support a toroidally dominated (or helical with a dominant toroidal component) magnetic field in relativistic jets. In particular, radio polarization observations reveal toroidally dominated magnetic fields on parsec scales \citep[e.g.,][]{2018Galax...6....5W}, optical polarization measurements indicate similar magnetic structures at smaller scales \citep[e.g.,][]{2008Natur.452..966M}, while recent X-ray polarization observations further suggest that the more upstream X-ray-emitting regions are likewise dominated by a helical magnetic field \citep[e.g.,][]{2023NatAs...7.1245D}. From the theoretical perspective, the toroidal magnetic field is not only widely adopted in studies of relativistic jets \citep[e.g.,][]{2019MNRAS.484.1378G}, but also needed to give an explanation for the observed flat radio spectra in standard conical jet models \citep{1979ApJ...232...34B,2012MNRAS.423..756P,2023MNRAS.526.5054L}. Accordingly, we adopt this observationally and theoretically motivated toroidal magnetic field evolution as a fiducial description, so that the radial evolution of magnetic field strength $B'(r_{\rm diss})$ can be estimated as
\begin{equation}\label{bevo}
B'(r_{\rm diss})=B'_{1\,\mathrm{pc}}(r_{\rm diss}/1\,\rm pc)^{-1}.
\end{equation}
Recent high-resolution observations of M\,87 jet \citep{2023A&A...673A.159R} suggest that the magnetic field evolution from parsec scales down to $\sim10\,R_{\rm S}$ can be approximately described by Eq.\,(\ref{bevo}), with the index lying approximately between $-0.7$ and $-1.2$. We therefore use this range to estimate the uncertainty associated with the assumed toroidal magnetic field evolution. Nevertheless, the innermost jet may remain poloidally dominated prior to the transition toward a toroidally dominated configuration, and the implications of such an alternative magnetic geometry are discussed further in Sect.\,\ref{sites}. Future improvements in constraining the magnetic field evolution on sub-parsec scales can be straightforwardly incorporated into the same framework to reassess where the condition $Y\gg1$ can be realized under alternative magnetic configurations.

\subsection{Klein-Nishina Suppression and Numerical Evaluation of $Y$}\label{ymin}
With $\Gamma(r_{\rm diss})$ and $B'(r_{\rm diss})$ estimated as described above, one can identify regions along the jet where the condition $Y \gg 1$ is satisfied. Observational constraints from the hard X-ray spectrum can be used for calibration, thereby constraining the minimum value of $Y$ (hereafter $Y_{\min}$) required to suppress synchrotron radiation from secondary pairs without overshooting the hard X-ray data. Since secondary pairs and external photons are both highly energetic, EC scattering occurs in the deep Klein--Nishina (KN) regime, where the EC to synchrotron flux ratio departs from the simple energy density scaling and becomes non-linear. To account for the KN suppression in the EC cooling of secondary pairs, we adopt the numerical correction proposed by \cite{2010NJPh...12c3044S}. Moreover, because the hard X-ray coverage and the secondary pairs spectral shape vary among sources, it is not straightforward to determine analytically which part of the hard X-ray spectrum most strongly constrains the synchrotron emission from secondary pairs. Therefore, a numerical evaluation with observational calibration is required to determine $Y_{\min}$.

The differential spectra of $\gamma$-ray photons, pairs, and neutrinos produced in the photopion ($p\gamma$) production and BH processes are obtained with analytical expressions developed in \cite{2008PhRvD..78c4013K}. The differential spectrum of pairs generated from the internal $\gamma \gamma$ annihilation is evaluated as well \citep{1983Afz....19..323A}. The spectra of these secondary pairs, produced through the above processes, are governed by the spectra of the parent relativistic protons and the external target photons. As described in Sect.~\ref{ext}, the spectral shape of external photon fields is fixed a priori; therefore, the relativistic proton spectrum must be assumed. In the numerical calculations, following standard expectations for diffusive shock acceleration \citep{1983RPPh...46..973D}, we adopt a proton spectral index of $-2$ and fix the proton Lorentz factor in the range $1 \le \gamma_{\rm p} \le 10^7$. The maximum proton Lorentz factor mainly affects the spectra peak of the secondary pairs emission, whereas the synchrotron emission in the X-ray band is produced predominantly by comparatively lower-energy pairs. Therefore, in the context of interpreting the $\sim1\,\mathrm{PeV}$ neutrino event, the choice $\gamma_{\rm p,\max}=10^7$ is reasonable and conservative. 

In the numerical evaluation of $Y_{\min}$, the value of $\delta$ at different jet position is derived from $\delta_{\rm C}$ using Eq.\,(\ref{profile}).
Then the proton injection luminosity is adjusted to match the observed neutrino flux, after which $B'$ is varied to scan $Y$ and obtain the corresponding secondary-pair emission spectra. This approach isolates the impact of $B'$ variations on $U'_{\rm B}$ and thus $Y$, while keeping the deep KN effect essentially fixed. In contrast, changing $\delta$ would complicate the interpretation of $Y_{\min}$ due to its effects on both synchrotron spectral features and KN effects. For each $Y$, we calculate the secondary pairs emission spectra and determine $Y_{\rm min}$ that remains consistent with the hard X-ray constraints.
\begin{figure}[htbp]
\subfigure{
\includegraphics[width=0.5\textwidth]{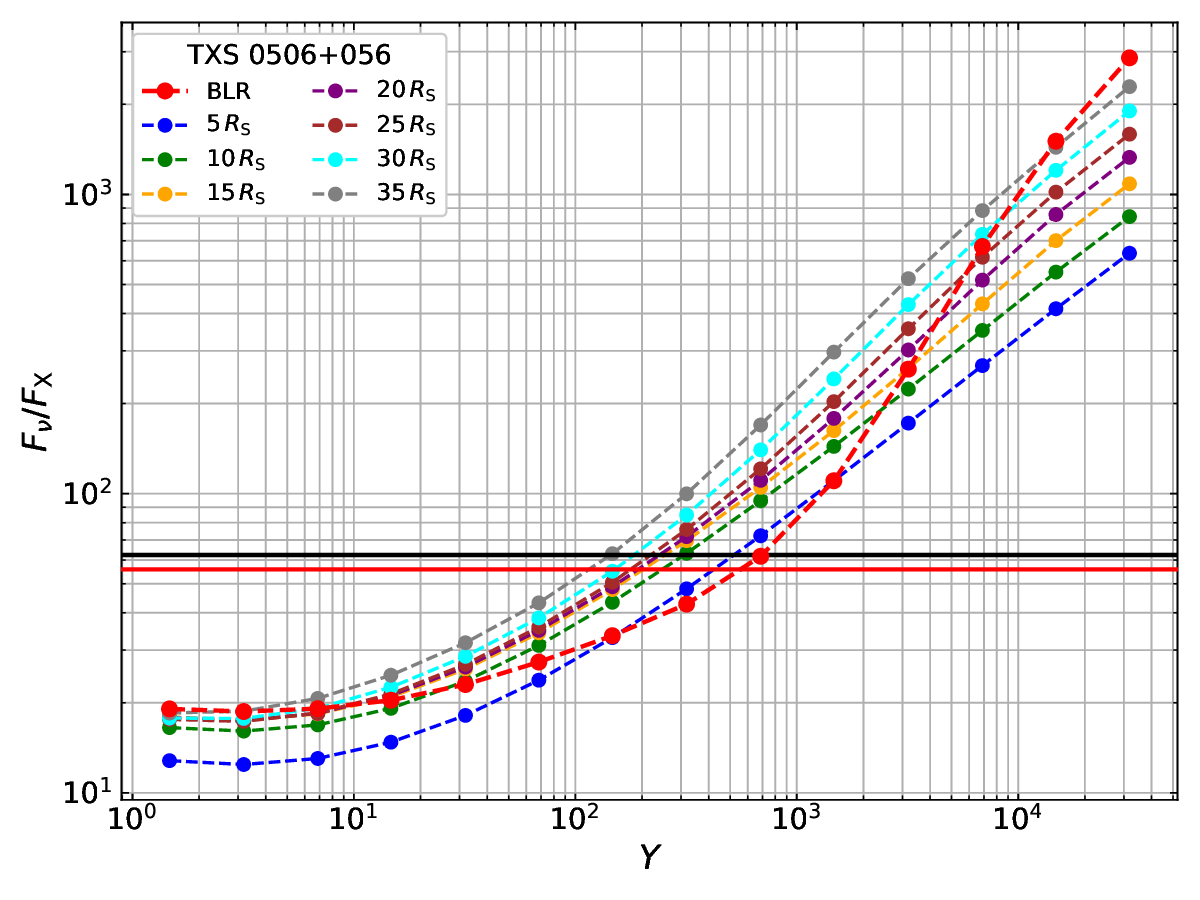}
}\hspace{-5mm}
\quad
\subfigure{
\includegraphics[width=0.5\textwidth]{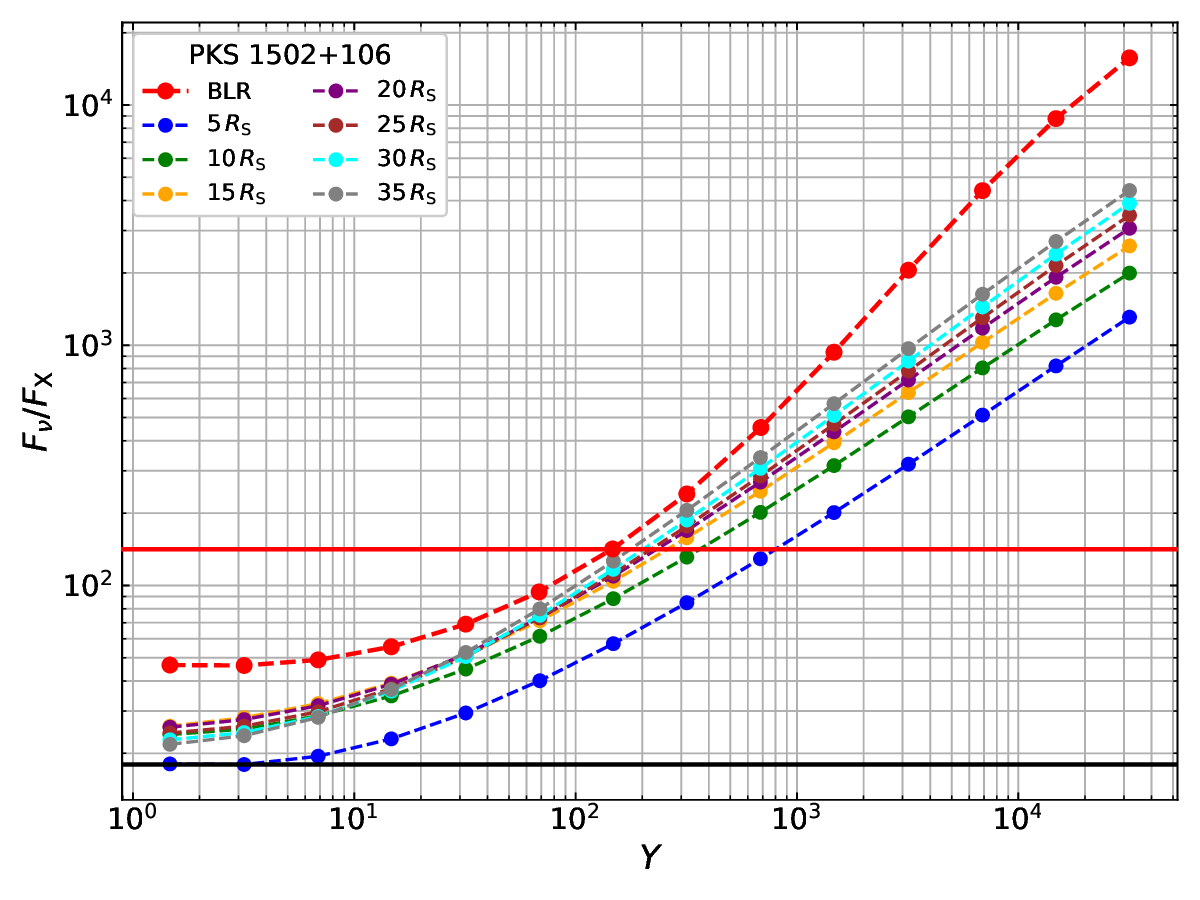}
}
\caption{Ratio of the observed neutrino flux (0.2--7.5\,PeV) to the integrated X-ray flux as a function of the energy density ratio between the external photon field and the magnetic field ($Y=U'_{\rm ext}/U'_{\rm B}$). The upper and lower panels display the results for TXS\,0506+056 and PKS\,1502+106, respectively. The external photon field is dominated by the AD within 5--35\,$R_{\rm S}$ and by the BLR beyond this range. Solid black and red horizontal lines represent the observational constraints derived from AD-dominated and BLR-dominated scenarios, respectively; the discrepancy between these horizontal lines arises from the different spectral shapes of secondary pair radiation, which necessitates constraints from distinct X-ray energy bands. Dashed lines in varying colors denote modeling results at different jet positions, as specified in the legend.
\label{Ymin}}
\end{figure}
\section{Application to Individual Sources} \label{sec:app}
In this section, we apply the framework developed above to two representative blazars, TXS\,0506+056 and PKS\,1502+106. They provide ideal case studies for our method for two reasons. First, both of them have VLBI radio measurements that independently constrain $B'$ and $\delta$. Second, they have quasi-simultaneous MWL SEDs, which reduce the ambiguity introduced by state-dependent variability.

\subsection{Determination of \texorpdfstring{$Y_{\min}$}{Ymin}}
Given that the observed energy $E_{\nu}^{\rm obs}$ of high-energy neutrinos associated with blazars typically falls within $0.1\text{--}1~\rm PeV$, the $\delta$-approximation yields an estimated external target photon energy of $E_{\rm ext}^{\rm AGN} \approx 10\text{--}1\,{\rm eV} (1/z) (0.1\text{--}1\,\rm PeV / \textit{E}_{\nu}^{\rm obs})$ in the AGN frame. This energy range corresponds to the optical--UV band, thereby excluding the DT as a primary source of target photons. Furthermore, to suppress the synchrotron emission from secondary pairs, it is preferable that EC scattering occurs in relative weak KN effect. Optical--UV photons originating from the AD and BLR satisfy this requirement, whereas the HC mainly emits the X-ray photons, where KN effect would be more severe (in particular, as shown in Fig.\,\ref{Urdiss}, $U'_{\rm HC}$ is always lower than $U'_{\rm AD}$; hence the HC can be safely excluded when determining $Y_{\rm min}$). Consequently, the determination of $Y_{\rm min}$ should primarily focus on locations near the AD and BLR, as these sites provide the most favorable conditions for neutrino production while minimizing synchrotron emission from secondary pairs.

Following Sect.\,\ref{ymin}, we calculate the ratio of neutrino flux to X-ray flux $F_{\nu}/F_{\rm X}$ as a function of $Y$ at different dissipation locations. The results are shown in Fig.\,\ref{Ymin} for TXS\,0506+056 and PKS\,1502+106, respectively. In both sources, $F_{\nu}/F_{\rm X}$ increases with $Y$ but with a non-linear scaling due to KN suppression. In the AD-dominated region (i.e., $<40\,R_{\rm S}$, where the AD photons fall within the required $\sim$1--10\,eV target-energy range; as shown in Fig.\,\ref{Urdiss}), increasing $r_{\rm diss}$ shifts the photon field of AD to lower energies in the comoving frame, alleviating KN effect and thus reducing the $Y$ required to suppress secondary pairs synchrotron emission. In contrast, in the BLR-dominated regime, external photons are boosted to higher energies, pushing the EC scattering deeper into the KN regime and weakening the EC efficiency, resulting in a systematically larger $Y_{\min}$. The horizontal lines give the observational lower limits on $F_{\nu}/F_{\rm X}$. Their intersections with the model curves define $Y_{\min}$, yielding $Y_{\min} \sim 1.5\times10^{2}$--$6\times10^{2}$ for TXS\,0506+056. For PKS\,1502+106, $Y_{\min} \gtrsim 3.2$ in the AD-dominated scenario (i.e., the condition is satisfied in essentially all cases) and $\sim 1.5\times10^{2}$ in the BLR-dominated scenario. This pronounced discrepancy arises from the distinct spectral shapes of the secondary pairs emission in the two scenarios, which results in markedly different observational constraints, as illustrated by the widely separated horizontal lines in the lower panel of Fig.\,\ref{Ymin} (see Appendix\,\ref{sec} for more details). 

\subsection{Locating the Potential Neutrino Production Sites with Radio Observations}\label{sites}

\begin{figure}[htbp]
\subfigure{
\includegraphics[width=0.5\textwidth]{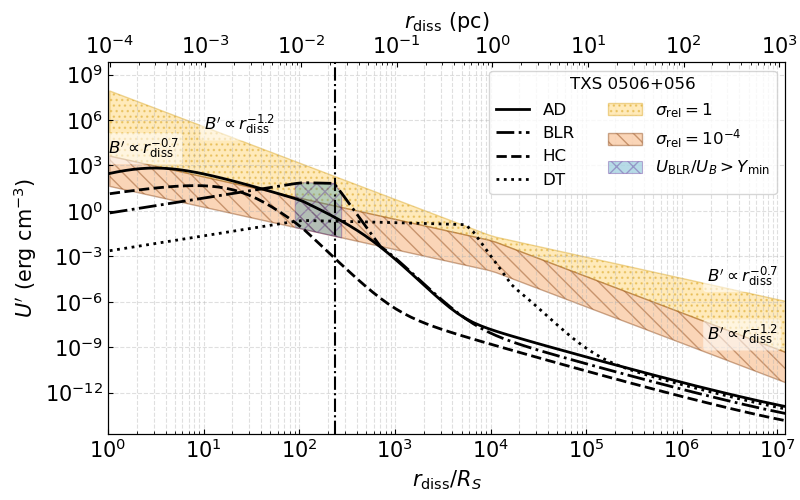}
}
\quad
\subfigure{
\includegraphics[width=0.5\textwidth]{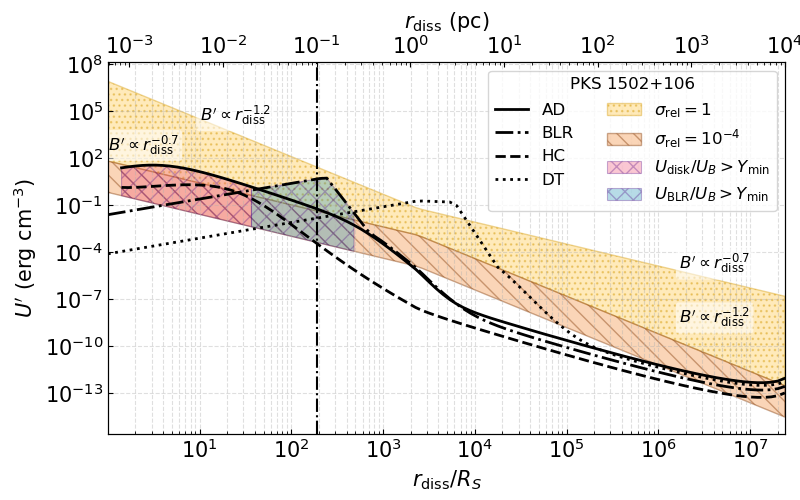}
}
\caption{Comoving energy densities $U'$ of external radiation fields and the magnetic field as a function of $r_{\mathrm{diss}}$ for TXS\,0506+056 (upper panel) and PKS\,1502+106 (lower panel). Black curves represent the energy densities of the AD (solid line), HC (dashed line), BLR (dash-dotted line), and DT (dotted line). The vertical black dash-dotted line represents the $R_{\mathrm{BLR}}$. The yellow and orange shaded bands illustrate the $U'_{\mathrm{B}}$ profiles assuming $\sigma_{\mathrm{rel}} = 1$ and $\sigma_{\mathrm{rel}} = 10^{-4}$, bounded by magnetic field scalings of $B' \propto r_{\mathrm{diss}}^{-0.7}$ and $B' \propto r_{\mathrm{diss}}^{-1.2}$. The pink and blue cross-hatched areas highlight the spatial regions satisfying $U'_{\mathrm{disk}}/U'_{\mathrm{B}} > Y_{\mathrm{min}}$ and $U'_{\mathrm{BLR}}/U'_{\mathrm{B}} > Y_{\mathrm{min}}$, respectively.
\label{Urdiss}}
\end{figure}

To locate potential neutrino production sites satisfying $Y > Y_{\mathrm{min}}$, we evaluate the radial profiles of $U'_{\rm ext}(r_{\rm diss})$ and $U'_{\rm B}(r_{\rm diss})$ using the method described in Sects.\,\ref{ext} and \ref{radio}. High-resolution radio observations provide key constraints on $\delta_{\rm C}$ and $B'^{\rm B,e}_{\rm 1\,pc}$. For TXS\,0506+056, VLBI measurements quasi-simultaneous with IC-170922A yield $\delta_{\rm C}\approx \Gamma_{\rm C}\geqslant13.58$ at the radio core, based on the equipartition brightness temperature assumption, and $B'^{\rm B,e}_{\rm 1\,pc}\simeq0.51$--$0.72\,$G \citep{2020ApJ...896...63L}. For PKS\,1502+106, the constraints are given by the archival MOJAVE program. Adopting a statistically constrained intrinsic brightness temperature of $T_{\rm b,int}=10^{10.609}\,{\rm K}$ (instead of the standard equipartition value) leads to a large Doppler factor and bulk Lorentz factor, i.e., $\delta_{\rm C}\approx65.2$ and $\Gamma_{\rm C}\approx34.9$ \citep{2021ApJ...923...67H}. The magnetic field strength is estimated to be $B'^{\rm B,e}_{\rm 1\,pc}\simeq0.69\pm0.52\,$G based on 2006 observations \citep{2012AA...545A.113P}.

\begin{table}[htbp]
\centering
\caption{Parameters adopted in Fig.~\ref{Urdiss}. Except for $r_{\rm tran}$, the parameters are adopted based on observational constraints.}
\label{Upara}
\begin{tabular*}{\columnwidth}{@{\extracolsep{\fill}}lll}
\toprule
Parameters & TXS\,0506+056 & PKS\,1502+106 \\
\midrule
$L_{\rm AD}$ (erg/s) & $5 \times 10^{44}$ [1] & $1 \times 10^{46}$ [2] \\
$M_{\rm BH}$ ($M_\odot$) & $10^{9}$ [1] & $10^{9.64}$ [3] \\
$\delta_{\rm C}$ & 30 & 65.2 [4] \\
$\Gamma_{\rm C}$ & 30 & 34.9 [4] \\
$r_{\rm C}$\,(pc) & 3.8 [5] & 8.19 [4] \\
$B'^{\rm B,e}_{1\,\rm pc}$\,(G) & 0.51--0.72 [5] & 0.17--1.21 [6] \\
$r_{\rm tran}$\,(pc) & 0.01 & 1 \\
\bottomrule
\end{tabular*}
\raggedright
\textbf{Notes.} [1] \cite{2019MNRAS.484L.104P}; [2] \cite{2021JCAP...10..082O}; [3] \cite{2011ApJS..194...45S}; [4] \cite{2021ApJ...923...67H}; [5] \cite{2020ApJ...896...63L}; [6] \cite{2012AA...545A.113P}. For TXS\,0506+056, $\delta_{\rm C}=30$ is adopted, which is consistent with the lower limit of 13.58 given by [5].
\end{table}

Using these observational constraints, the resulting $U'_{\rm ext}(r_{\rm diss})$ and $U'_{\rm B}(r_{\rm diss})$ are shown in Fig.\,\ref{Urdiss}, and the corresponding parameters are given in Table\,\ref{Upara}. Adopting $\sigma_{\rm rel}=10^{-4}$ as a typical value in hadronic models, both TXS\,0506+056 and PKS\,1502+106 exhibit viable regions with $Y>Y_{\min}$, though with different extents. For TXS\,0506+056, it allows only a limited region. Even assuming a relatively small transition distance $r_{\rm tran}=0.01\,$pc (as shown in the upper panel of Fig.\,\ref{Urdiss}), a viable parameter space appears only for sufficiently strong beaming, i.e., $\delta_{\rm C}\gtrsim16$. For $\delta_{\rm C}=30$, the allowed region is confined to a limited portion near the BLR. In contrast, assuming $r_{\mathrm{tran}} = 1\,\mathrm{pc}$ (as shown in the lower panel of Fig.\,\ref{Urdiss}), PKS\,1502+106 features a broad allowed region, with $Y>Y_{\min}$ satisfied throughout most radii within $\sim3\,R_{\rm BLR}$ due to its large $\Gamma_{\rm C}$ boosting $U'_{\rm ext}$. Also, the relatively low lower limit of $B'^{\rm B,e}_{1\,\rm pc}$ than that of TXS\,0505+056 is another reason that caused a wider parameter space. In addition, the larger $L_{\rm AD}$ of PKS\,1502+106 shifts the characteristic BLR radius outward ($R_{\rm BLR}\propto L_{\rm AD}^{1/2}$). Although this scaling does not increase the external radiation energy density inside the BLR, it enlarges the BLR-dominated region in absolute distance, where $U'_{\rm B}$ has already declined, and therefore can further enlarge the radial range with large $Y$. More generally, viable parameter space exists for much larger $r_{\rm tran}$ as well. However, the jet energy budget imposes a further constraint on this parameter space. Since $U'_{\rm BLR}$ drops rapidly at larger radii, sustaining a neutrino rate of $\sim 1 \rm \ yr^{-1}$ may require super-Eddington jet power. While transient super-Eddington flares are possible, a physically plausible sustained source should remain sub-Eddington, effectively confining the production site to the BLR vicinity. As discussed in Sect.\,\ref{radio}, the toroidally dominated magnetic field adopted in this work should be regarded as a fiducial description. An alternative possibility is that the innermost jet may be highly magnetized and dominated by a poloidal magnetic field near the SMBH, where the jet is still accelerating and retains a parabolic geometry \citep{2017Galax...5...11G,2026ApJ...996L..22P}. In such a configuration, the magnetic field scale as $B'\propto r_{\rm diss}^{-2}$, substantially enhancing $U'_{\rm B}$ in the inner jet. This makes the condition $Y > Y_{\min}$ harder to satisfy, implying that the AD-dominated scenario may not be generically viable and is instead highly sensitive to the jet's magnetic structure. Therefore, the fiducial toroidal magnetic field assumption adopted here should be regarded as a relatively optimistic case for realizing the large-$Y$ condition at the smallest jet scales. Nevertheless, although the detailed extent of the allowed parameter space depends on the magnetic field evolution, the BLR vicinity remains a more robust and conservative site where $Y>Y_{\min}$ can be achieved.
We note that the quantitative value of $Y$ also depends on the adopted value of $\sigma_{\rm rel}$. Since $B'\propto\sigma_{\rm rel}^{1/4}$, the magnetic energy density scales as $U'_{\rm B}\propto\sigma_{\rm rel}^{1/2}$, implying $Y\propto\sigma_{\rm rel}^{-1/2}$. Therefore, increasing $\sigma_{\rm rel}$ by one or two orders of magnitude (i.e., to $10^{-3}$ or $10^{-2}$) would reduce $Y$ by factors of $\sim3$ and $\sim10$, respectively, thereby shrinking the parameter space satisfying $Y>Y_{\rm min}$. This effect would be more pronounced for objects such as TXS\,0506+056, where only a limited parameter space is available under our fiducial assumption (i.e., $\sigma_{\rm rel}=10^{-4}$), whereas PKS\,1502+106 would remain comparatively less affected owing to its much larger $Y$. Conversely, adopting a smaller $\sigma_{\rm rel}$ would enlarge the allowed parameter space. Nevertheless, our qualitative conclusion remains unchanged, and efficient neutrino production is still favored in the vicinity of the BLR.

Overall, the requirement $Y>Y_{\min}$ can be met either by (i) a small transition distance $r_{\rm tran}$, which makes the jet reach a large $\Gamma$ on sub-parsec scales and thereby enhances $U'_{\rm ext}$, or (ii) a large $\Gamma_{\rm C}$, which also boosts $U'_{\rm ext}$ even when the transition occurs at larger radii. In this context, TXS\,0506+056 exemplifies the ``small-$r_{\rm tran}$'' scenario, where $r_{\rm tran}<R_{\rm BLR}$ is required to achieve sufficiently large $\Gamma$ near the BLR. By contrast, PKS\,1502+106 represents the ``large-$\Gamma_{\rm C}$'' scenario, in which $Y>Y_{\min}$ is satisfied over a broad region even with $r_{\rm tran}\gtrsim R_{\rm BLR}$. These two scenarios, although seemingly distinct, reflect a same physical requirement, i.e., a sufficiently large $\Gamma$ in the region near or within the BLR. It simultaneously enhances $U'_{\rm ext}$, thereby increasing the neutrino production efficiency and alleviating the energy budget by reducing the required power in relativistic protons below the Eddington luminosity, and leads to a large $Y$, under which the synchrotron emission from secondary pairs is naturally suppressed, ensuring consistency with the X-ray constraints. 

However, observations suggest that the above requirements are expected to be met only in exceptional cases. Radio studies of nearby jetted AGN find the transition from parabolic to conical geometry typically occurs at $\sim 1$--$10\,{\rm pc}$, i.e., generally beyond $R_{\rm BLR}$ \citep{2020MNRAS.495.3576K}, suggesting that AGN jets with $r_{\rm tran}< R_{\rm BLR}$ are uncommon. It is worth noting that this observed transition scale of $1$--$10\,$pc is broadly consistent with the typical Bondi radius \citep{2012ApJ...745L..28A,2020MNRAS.495.3576K}. In this context, while the $r_{\rm tran}$ required in the ``large-$\Gamma_{\rm C}$'' scenario can be comparable to the Bondi radius, the requirement of $r_{\rm tran} \sim 0.01\,$pc in the ``small-$r_{\rm tran}$'' scenario demands a geometry transition at a location significantly smaller than this characteristic physical scale. Notably, none of the ten nearby jetted AGNs in the sample of \citet{2020MNRAS.495.3576K}, which exhibit these standard transition scales, are included in the IceCat-1 catalog \citep{2023ApJS..269...25A}. This absence further supports the idea that a large $r_{\rm tran}$ is generally unfavorable for neutrino production, unless compensated by an exceptionally large $\Gamma_{\rm C}$. However, the ``large-$\Gamma_{\rm C}$'' scenario itself represents an extreme case in the blazar population, as blazars with $\Gamma_{\rm C}$ comparable to PKS\,1502+106 constitute only a small fraction of the known population \citep[e.g.,][]{2018ApJ...866..137L}. Consequently, whether a jet undergoes an uncommon early geometry transition or possesses extreme relativistic beaming, the conditions required for efficient neutrino emission remain difficult to fulfill. This naturally explains why only a limited subset of blazars has been identified as high-energy neutrino sources at present.

\subsection{SED Modeling and Implications for the Neutrino Events}
Building on the results above, the potential neutrino production regions for both TXS\,0506+056 and PKS\,1502+106 are likely situated near or within the BLR, where $Y$ typically reaches values of $\sim 10^2$--$10^3$. However, their SEDs show that the ratio of the high-energy peak to the low-energy peak flux generally does not exceed $\sim 10$. This discrepancy implies that a single-zone scenario cannot simultaneously account for both the neutrino emission and the broadband SED, and instead requires the introduction of an additional leptonic dissipation region.

\begin{figure*}[htbp]
\subfigure{
\includegraphics[width=0.5\textwidth]{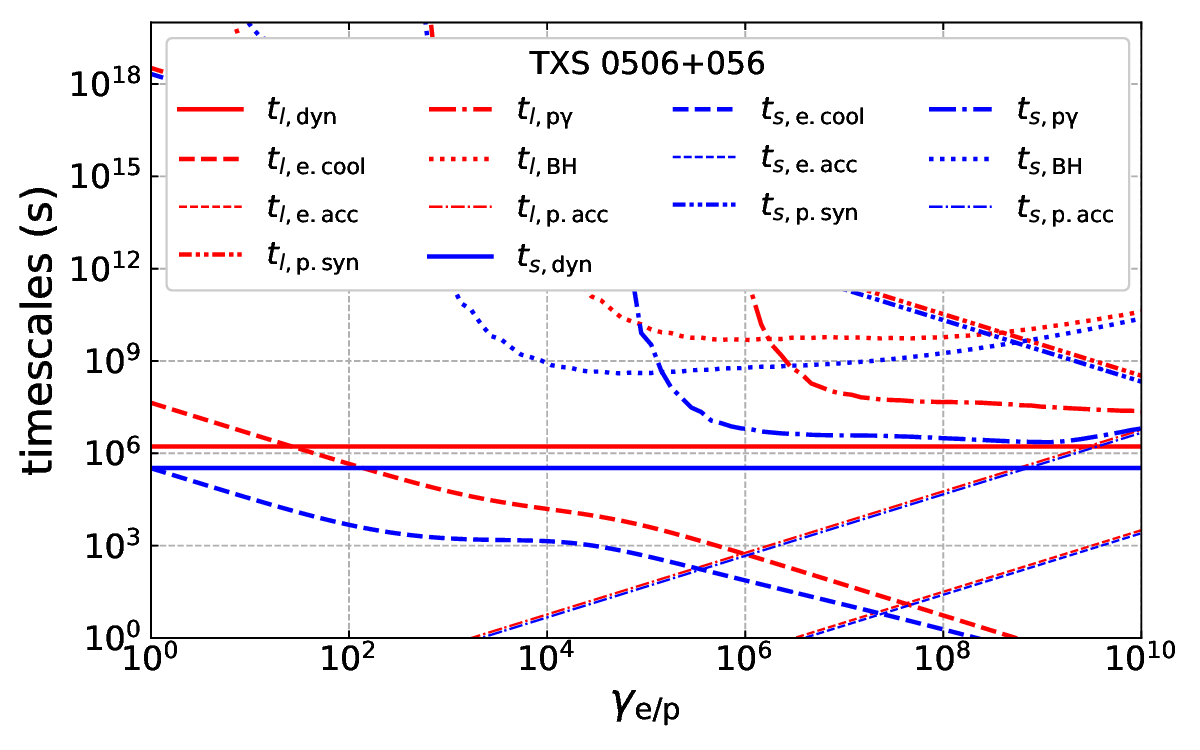}
}
\quad
\subfigure{
\includegraphics[width=0.5\textwidth]{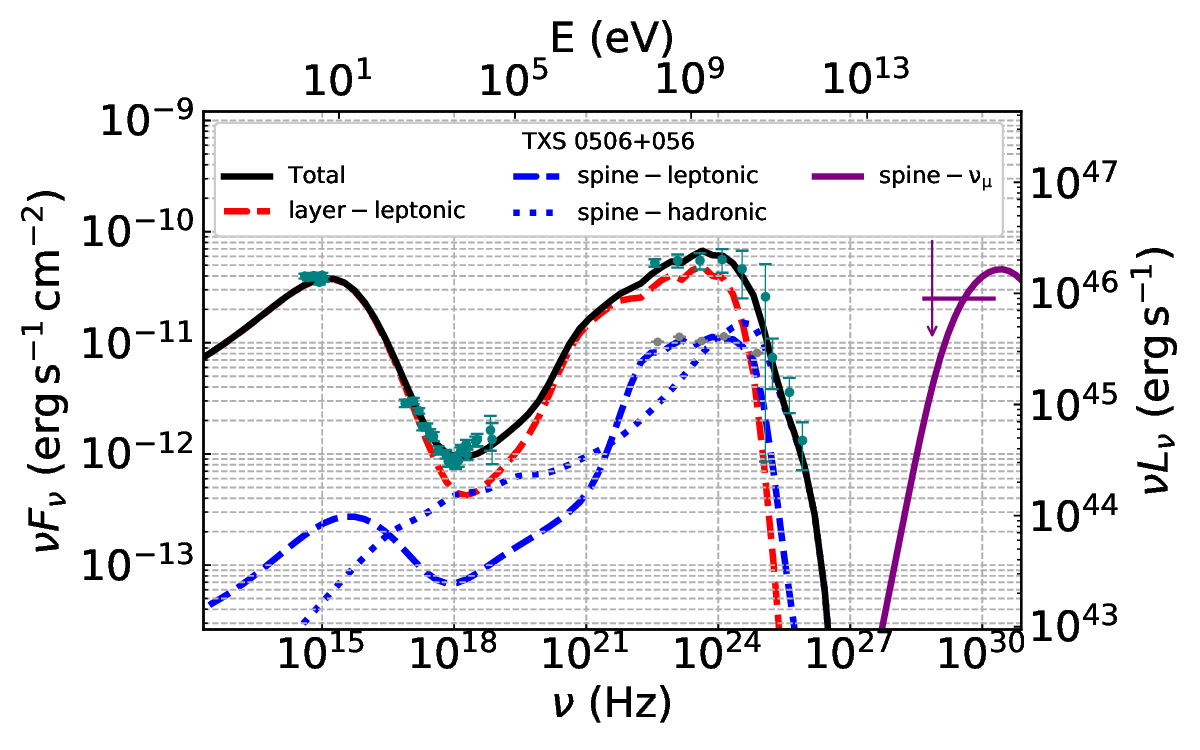}
}
\quad
\subfigure{
\includegraphics[width=0.5\textwidth]{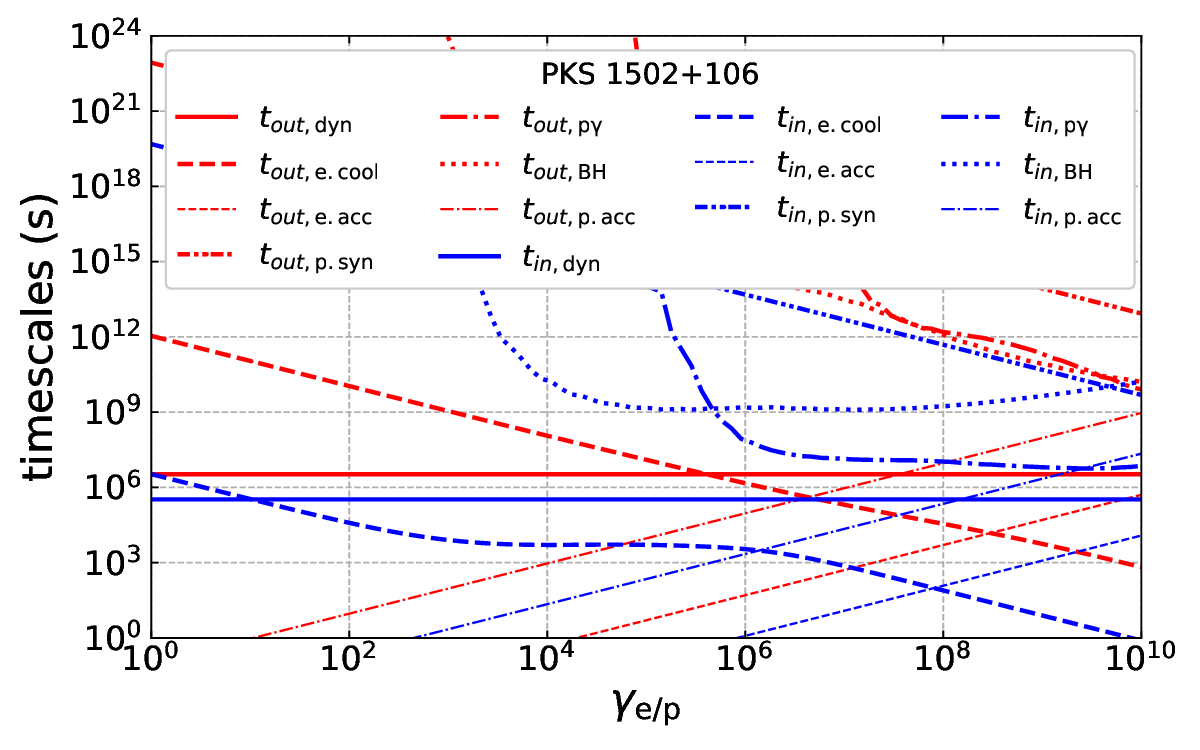}
}
\quad
\subfigure{
\includegraphics[width=0.5\textwidth]{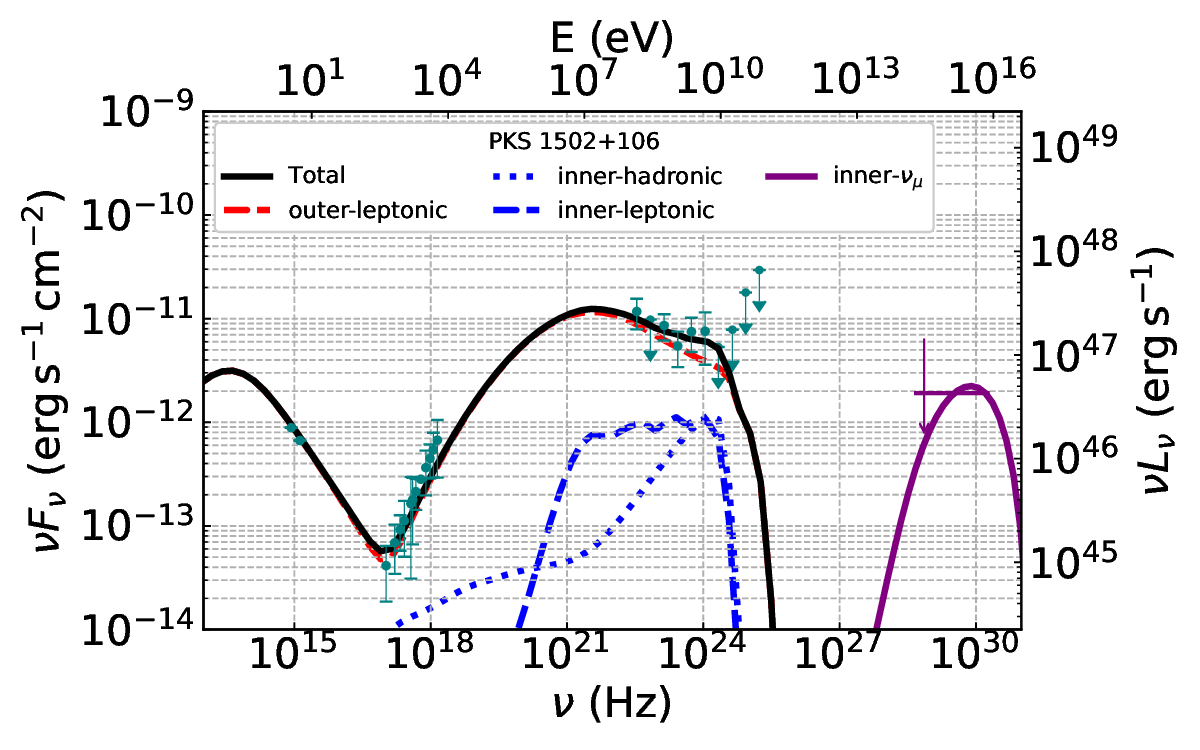}
}
\caption{The left panels show the particle acceleration, cooling and dynamical timescales for the spine--layer model of TXS\,0506+056 (\textbf{upper left}) and the inner--outer blob model of PKS\,1502+106 (\textbf{lower left}). In each panel, the blue curves correspond to the spine or inner blob, while the red curves correspond to the layer or outer blob. Thick solid, dashed, dotted, dot-dashed, and double dot-dashed curves represent the dynamical timescales, the electron cooling timescales (including synchrotron, SSC, and EC cooling), BH cooling timescales, $p\gamma$ cooling timescales, and proton synchrotron cooling timescales, respectively. Thin dashed, and dot-dashed curves represent the diffuse shock acceleration timescales $t_{\rm e/p.acc}\simeq20\gamma_{\rm e/p}m_{\rm e/p}c/3eB'$ for electrons and protons under Bohm limit \citep{2007Ap&SS.309..119R}. The right panels show the predicted multimessenger emissions of TXS\,0506+056 (\textbf{upper right}) in the spine--layer blob model and PKS\,1502+106 (\textbf{lower right}) in the inner--outer blob model. The teal data points show quasi-simultaneous SEDs of TXS\,0506+056 and PKS\,1502+106 taken from \cite{2018Sci...361.1378I} and \cite{2020ApJ...893..162F}, respectively. The grey data points are the historical flux listed in the Fermi-LAT 3FGL catalog \citep{2015ApJS..218...23A}. The dashed red curves represent the leptonic emission from primary electrons in the layer or outer blob. The dashed and dotted blue curves represent the leptonic emission of primary electrons and secondary pairs in the spine or inner blob, respectively. The solid black curves are the total emission from both dissipation regions. The purple curves represent the muon neutrino energy spectra from the spine or inner blob. The solid purple lines show the upper limits on the $\nu_\mu+\bar{\nu}_\mu$ neutrino flux to produce one event over a period of one year, assuming a neutrino spectrum of $E^{-2}$ in the range of 0.2--7.5\,PeV.
\label{SED}}
\end{figure*}

For the association between TXS\,0506+056 and IC-170922A, pronounced MWL variability from optical to TeV bands was observed around the neutrino arrival time \citep{2018Sci...361.1378I}, suggesting that the neutrino production is closely linked to the MWL activity. This temporal coincidence indicates that the neutrino-emitting region and the broadband radiation zone are likely physically connected. Here we consider a spine--layer jet model \citep{2005A&A...432..401G}, in which a fast spine is surrounded by a slow layer. Limb-brightened structures observed on $\sim100\,\rm pc$ scales \citep{2020A&A...633L...1R} provide evidence for a structured jet with transverse velocity stratification in TXS\,0506+056, suggesting that such a structure may also extend to sub-parsec scales relevant to neutrino emission. In this scenario, the spine region, characterized by a larger bulk Lorentz factor, experiences a strongly boosted external radiation field from the BLR, leading to efficient neutrino production. In contrast, the surrounding layer, moving with a lower bulk Lorentz factor, sees a significantly reduced external photon energy density. In the modeling, the evaluation of synchrotron, synchrotron self-Compton (SSC), and EC emissions (including photons from the AD, HC, BLR, DT, and the other component) from primary electrons and secondary pairs in both the spine and layer follows \cite{2005A&A...432..401G}. For the hadronic processes in the spine, we calculate the differential spectra of $\gamma$-ray photons from neutral pion decays, pairs, and neutrinos following \cite{2008PhRvD..78c4013K}. Besides, the pairs coming from internal $\gamma\gamma$ pair production are also calculated following \cite{1983Afz....19..323A}. Although relativistic protons may also be present in the layer, its smaller bulk Lorentz factor leads to a substantially lower $U'_{\rm ext}$, thereby making the resulting hadronic emission negligible. For both the spine and layer, relativistic electrons and protons are injected with prescribed injection spectra. The steady-state distributions of these primary particles are obtained by solving the continuity equations including particle injection, cooling, and escape. The corresponding continuity equations for secondary pairs (including generation, cooling, and escape) and photons (including generation, synchrotron self-absorption and internal $\gamma\gamma$ absorption) are solved simultaneously. These coupled continuity equations are solved iteratively until a self-consistent steady-state solution is reached. The detailed numerical approach follows \cite{2025EPJC...85..779X}.
The absorption of $\gamma$-ray escaping from the jet will be absorbed by the extragalactic background light (EBL). We account for this attenuation in the GeV--TeV band using the EBL model of \cite{2011MNRAS.410.2556D}.

\begin{table}[htbp]
\centering
\begin{threeparttable}
\caption{Parameters of the spine-layer jet model for TXS\,0506+056.}
\label{slpara}
\begin{tabular*}{\columnwidth}{@{\extracolsep{\fill}}lll}
\toprule
\textbf{Parameter} & \textbf{Spine} & \textbf{Layer} \\
\midrule

$R'$ (cm) & $6\times10^{16}$ & $3\times10^{17}$ \\
$L'$ (cm) & $1\times10^{16}$ & $5\times10^{16}$ \\
$B'$ (G) & $1.5$ & $1.1$ \\
$\Gamma \approx \delta$ & $48.3$ & $4$ \\
$r_{\rm diss}$ (pc) & $2.2\times10^{-2}$ & $2.2\times10^{-2}$\\
\midrule

\multicolumn{3}{c}{\textit{Injected primary electron population}} \\
$n_1$ & $2$ & $1.7$ \\
$n_2$ & $4$ & $4.5$ \\
$\gamma_{\rm e,min}$ & $5\times10^{1}$ & $1\times10^{2}$ \\
$\gamma_{\rm e,b}$ & $5\times10^{3}$ & $1.5\times10^{4}$ \\
$\gamma_{\rm e,max}$ & $1\times10^{7}$ & $1\times10^{7}$ \\
$L'_{\rm e,inj}$ (erg s$^{-1}$) & $1\times10^{40}$ & $3\times10^{45}$ \\
\midrule

\multicolumn{3}{c}{\textit{Injected primary proton population}} \\
$n_p$ & $2$ & --- \\
$\gamma_{\rm p,min}$ & $1$ & --- \\
$\gamma_{\rm p,max}$ & $6\times10^{7}$ & --- \\
$L'_{\rm p,inj}$ (erg s$^{-1}$) & $1.3\times10^{42}$ & --- \\
\bottomrule
\end{tabular*}

\begin{tablenotes}
\item \footnotesize \textbf{Notes.} 
$R'$ and $L'$ denote the radius and width of the emitting region. Injected electron populations follow broken power-law distributions with indices $n_1, n_2$, Lorentz factors $\gamma_{\rm e,min}, \gamma_{\rm e,b}, \gamma_{\rm e,max}$, and injection luminosities $L'_{\rm e,inj}$. Injected proton populations follow a power-law distribution with index $n_p$, Lorentz factors $\gamma_{\rm p,min}, \gamma_{\rm p,max}$, and injection luminosities $L'_{\rm p,inj}$. As shown in the upper left panel of Fig.\,\ref{SED}, $\gamma_{\rm e/p,max}$ adopted here is below the value at which the acceleration timescale intersects the cooling or dynamical timescale. The spine bulk Lorentz factor $\Gamma_s$ is inferred from $\Gamma_{\rm C}$ in Table\,\ref{Upara} using Eq.\,(\ref{profile}). The spine magnetic field $B'_s$ is inferred from $B'^{\rm B,e}_{1\,\rm pc}$ in Table\,\ref{Upara} by first converting $B'^{\rm B,e}_{1\,\rm pc}$ to $B'_{\rm 1\,pc}$ using Eq.\,(\ref{b1pc}) with $\sigma_{\rm rel}=10^{-4}$, and then applying Eq.\,(\ref{bevo}), whose index is allowed to vary between $-0.7$ and $-1.2$. The remaining parameters are obtained from the SED modeling. Proton parameters for the layer are omitted because their hadronic contribution is negligible.
\end{tablenotes}
\end{threeparttable}
\end{table}

The upper left and right panels of Fig.\,\ref{SED} show the timescales and modeling result of spine--layer model, and the corresponding parameters are listed in Table\,\ref{slpara}. Given that the detection of VHE photons implies that the dissipation region cannot be located within the BLR due to severe $\gamma\gamma$ absorption, we place the neutrino-emitting region just outside the BLR at $r_{\rm diss} = 0.022\,\rm pc$. For the spine, its $\Gamma_s$ and $B'_s$ (`$s$' refers to spine) are inferred by the radio-constrained profiles shown in Fig.\,\ref{Urdiss}, and the adopted $R'_s$ is consistent with the day-scale variability suggested by MAGIC \citep{2018ApJ...863L..10A}. The jet power is estimated as $P_{\rm jet}=\Gamma^2(L'_{\rm e, inj} + L'_{\rm p, inj}) + L_{\rm B}\approx 0.7L_{\rm Edd}$, where $L_{\rm B}=\pi R^2c\Gamma^2U'_{\rm B}$ and $L_{\rm Edd}$ is the Eddington luminosity. The cooling timescales shown in the upper left panel of Fig.~\ref{SED} demonstrate that the hadronic contribution from the layer is negligible. Its much smaller $\Gamma_l$ results in both less efficient proton cooling and much weaker Doppler boosting, such that reproducing a neutrino flux comparable to that of the spine would require a proton injection luminosity about six orders of magnitude larger. We therefore neglect the layer hadronic emission and omit the corresponding proton parameters from Table~\ref{slpara}.
Our modeling adopts an alternative strategy compared to previous spine--layer studies \citep[e.g.,][]{2018ApJ...863L..10A}. While conventional approaches typically assume the spine dominates the broadband SED and the layer serves primarily as an external photon seed source, our strategy is dictated by the extreme cooling environment near the BLR. In this position, the spine inevitably reaches $Y_s \gg 1$, making it difficult to reconcile a high neutrino flux with the observed SED peak ratios. Consequently, we assign the primary contribution of the broadband SED to the layer, where $Y_l \sim 1$ (`$l$' refers to layer) allows for a suppressed Compton dominance. This requirement also constrains the layer parameters. With the dissipation region located just outside the BLR, the external radiation field is still dominated by BLR photons, yielding a comoving energy density of $U'_{\rm BLR} \approx \Gamma_l^2 \times 0.028\,\rm erg\,cm^{-3}$. For a typical bulk Lorentz factor of $\Gamma_l \sim 2$--$4$, maintaining $Y_l \sim 1$ requires the magnetic and external radiation energy densities to be comparable, thereby restricting the allowed range of $R'_l$ and $B'_l$. Under these conditions, the model remains physically self-consistent and naturally explains the observed temporal coincidence between the neutrino event and MWL flares, as both zones are co-spatial and dynamically coupled, and yields a neutrino detection rate of $\sim 1\,{\rm yr^{-1}}$ in the range of 0.2--7.5 PeV, consistent with the IceCube observation.

Interestingly, while this manuscript was under review, two independent VLBI studies of TXS\,0506+056 reported observational evidence for a transverse spine--layer structure on parsec scales \citep{2026arXiv260627435E, 2026arXiv260627430K}. Both works find that the radio emission is dominated by the slower layer rather than the fast spine, consistent with the physical picture adopted in our model in which the layer dominates the MWL emission while the spine primarily produces neutrinos. Furthermore, the VLBI observational jet properties, including the characteristic layer thickness and the bulk Lorentz factors of the spine and layer \citep{2026arXiv260627435E, 2026arXiv260627430K}, are broadly consistent with the parsec scale extrapolations of our sub-parsec modeling parameters. Although VLBI observations probe parsec scale rather than sub-parsec regions, they nevertheless provide independent observational support for the spine-layer jet structure employed in this work and strengthen the physical plausibility of our interpretation.

\begin{table}[htbp]
\centering
\begin{threeparttable}
\caption{Parameters of the inner-outer blob model for PKS\,1502+106.}
\label{iopara}
\begin{tabular*}{\columnwidth}{@{\extracolsep{\fill}}lll}
\toprule
\textbf{Parameter} & \textbf{Inner Blob} & \textbf{Outer Blob} \\
\midrule

$R'$ (cm) & $1\times10^{16}$ & $1\times10^{17}$ \\
$B'$ (G) & $0.32$ & $7.8\times10^{-3}$ \\
$\delta$ & $29.6$ & $62.9$ \\
$\Gamma$ & $14.8$ & $33.1$ \\
$r_{\rm diss}$ (pc) & $0.1$ & $10$ \\
\midrule

\multicolumn{3}{c}{\textit{Injected primary electron population}} \\
$n_1$ & 2 & $1.4$ \\
$n_2$ & 4 & $4.3$ \\
$\gamma_{\rm e,min}$ & $5\times10^{1}$ & $5\times10^{2}$ \\
$\gamma_{\rm e,b}$ & $7\times10^{3}$ & $7\times10^{3}$ \\
$\gamma_{\rm e,max}$ & $1\times10^{7}$ & $3\times10^{5}$ \\
$L'_{\rm e,inj}$ (erg s$^{-1}$) & $5\times10^{41}$ & $1.8\times10^{43}$ \\
\midrule

\multicolumn{3}{c}{\textit{Injected primary proton population}} \\
$n_p$ & $2$ & --- \\
$\gamma_{\rm p,min}$ & $1$ & --- \\
$\gamma_{\rm p,max}$ & $1.7\times10^{7}$ & --- \\
$L'_{\rm p,inj}$ (erg s$^{-1}$) & $1.7\times10^{44}$ & --- \\
\bottomrule
\end{tabular*}
\begin{tablenotes}
\item \footnotesize \textbf{Notes.} The injected electron population follows a broken power-law distribution with indices $p_1$ and $p_2$, minimum, break, and maximum electron Lorentz factors $\gamma_{\rm e,min}$, $\gamma_{\rm e,b}$, and $\gamma_{\rm e,max}$, and injection luminosity $L'_{\rm e,inj}$. 
The injected proton population follows a power-law distribution with index $n_p$, minimum, and maximum proton Lorentz factors $\gamma_{\rm p,min}$ and $\gamma_{\rm p,max}$, and injection luminosity $L'_{\rm p,inj}$. As shown in the lower left panel of Fig.\,\ref{SED}, $\gamma_{\rm e/p,max}$ adopted here is below the value at which the acceleration timescale intersects the cooling or dynamical timescale.
The $\Gamma$($\delta$), and $B'$ of both blobs are inferred from the radio-constrained values listed in Table\,\ref{Upara} following the same procedure as described in the notes of Table\,\ref{slpara}. The remaining parameters are obtained through the SED modeling. Proton parameters for the outer blob are omitted because the hadronic emission is extremely weak.
\end{tablenotes}
\end{threeparttable}
\end{table}

For the association between PKS\,1502+106 and IC-190730A, the source was in a low state at the time of the neutrino arrival \citep{2020ApJ...893..162F}. Here, we employ the ``inner--outer blob'' model proposed in our previous work \citep{2019PhRvD..99f3008L,2019ApJ...886...23X}. In this picture, the inner blob (with parameters denoted by the subscript `in') represents the neutrino production zone located around the BLR, where $\Gamma_{\rm in}$, $\delta_{\rm in}$ and $B'_{\rm in}$ are determined by the radio-constrained profiles shown in Fig.\,\ref{Urdiss}. In contrast, the outer blob (denoted by `out') is responsible for the primary leptonic emissions that reproduce the SEDs. Its location can be inferred by identifying the region in Fig.\,\ref{Urdiss} where $Y$ matches the ratio of the high-energy to low-energy peak fluxes. Then, the corresponding $\Gamma_{\rm out}$, $\delta_{\rm out}$ and $B'_{\rm out}$ can be self-consistently inferred, providing an observationally-grounded basis for the subsequent SED modeling. Furthermore, both primary relativistic electrons and protons are injected in the inner blob. To remain consistent with the absence of a coincident electromagnetic flare, the electron injection luminosity is taken to be smaller than that of the protons. By contrast, hadronic emission from the outer blob is expected to be very weak because the proton cooling timescale is typically much longer than the dynamical timescale under the relatively weak target photon fields \citep{2019ApJ...886...23X}. Apart from the different geometry, the same self-consistent particle transport calculation is adopted as for the spine--layer model. The distributions of the injected primary particles, secondary pairs, and photons are evolved by solving the coupled continuity equations until a steady-state solution is reached.

The timescales and modeling result of the inner--outer model are shown in the lower right and left panels of Fig.\,\ref{SED}, and the corresponding parameters are listed in Table\,\ref{iopara}. The model reproduces the broadband SED with the outer blob, while the inner blob accounts for the neutrino production. The dashed blue curves in the lower right panel of Fig.\,\ref{SED} show that, with an electron injection luminosity about 300 times lower than that of the protons, the leptonic emission from the inner blob remains well below the observed SED. The outer blob is located at $\sim$10\,pc, where the external photon fields from the AD, HC, and BLR have all become very weak. Consequently, cooling of $p\gamma$ and BH interactions is highly inefficient, as confirmed by the cooling timescales shown in the lower left panel of Fig.\,\ref{SED}, where the proton cooling timescale is much longer than those in the inner blob. The resulting hadronic emission is several orders of magnitude below the dominant SED components and is therefore not visible. The corresponding proton parameters are therefore omitted from Table~\ref{iopara}, as they have no practical influence on the model predictions.
Consistent with the jet power constraints discussed in Sect.\,\ref{sites}, the total jet power is estimated to be $P_{\rm jet} \approx 0.1\,L_{\rm Edd}$. This sub-Eddington requirement justifies our placement of the inner blob near the BLR, as more distant sites would demand an unsustainable energy budget to produce the observed neutrino flux. For the adopted parameters, the inner region yield a neutrino detection rate of $\sim 1\,\rm yr^{-1}$ at the range of 0.2--7.5\,PeV, consistent with the IceCube observation. The inferred location of the outer blob at $10\,\rm pc$ corresponds to a region where the external radiation field is sufficiently weak, leading to $Y \sim 1$ that consistent with the observed SED. Meanwhile, the inner blob, residing close to the BLR, provides a dense photon field that ensures efficient neutrino production without violating the hard X-ray constraints. This separation of emission zones naturally explains the absence of a prominent MWL flare during the neutrino event, as the dominant electromagnetic emission arises from the outer region.

Overall, the choice of geometry for each source is driven by both observational and structural considerations. Observationally, the coincident MWL flare during IC-170922A favors a co-spatial structure (spine-layer) for TXS\,0506+056, while the low state of PKS\,1502+106 during IC-190730A favors a spatially separated structure (inner-outer). These choices are also consistent with the underlying jet acceleration profile: with $r_{\rm tran}\ll R_{\rm BLR}$ the jet reaches its terminal $\Gamma$ before crossing the BLR, making transverse stratification natural; with $r_{\rm tran}\gtrsim R_{\rm BLR}$ the jet is still accelerating, so longitudinal segmentation arises naturally from the radial $\Gamma$ gradient. In a general structured-jet picture both can coexist, and the two sources represent limiting cases.

\section{General Application to Blazar Population}\label{sec:popu}
Following the case studies presented above, the method developed in Sect.\,\ref{sec:method} can be naturally extended to a blazar sample, enabling a preliminary screening of neutrino candidates by combining radio constraints with X-ray observations.

Here we cross-match the MOJAVE catalog and identify a sample of 82 sources (see Table\,\ref{tab:mojave_sample} in Appendix\,\ref{sample}) for which independent constraints on both $B'^{\rm B,e}_{\rm 1\,pc}$ and $\Gamma_{\rm C}$ are available \citep{2012AA...545A.113P,2021ApJ...923...67H}. To enable a uniform comparison across the sample, we extrapolate $\Gamma_{\rm C}$ to a distance of 0.06\,pc, lying between the characteristic BLR scales of TXS\,0506+056 and PKS\,1502+106, using Eq.\,(\ref{profile}). We then use the ratio $\Gamma_{\rm 0.06\,pc}/B'^{\rm B,e}_{\rm 1\,pc}$ as a proxy for $Y = U'_{\rm ext}/U'_{\rm B}$, capturing the relative strength of external photon fields to magnetic fields. We compare the values of $\Gamma_{\rm 0.06\,pc}/B'^{\rm B,e}_{\rm 1\,pc}$ for these sources with that of PKS\,1502+106, an extreme beaming but possible candidate for neutrino emission in the MOJAVE sample. This comparison provides a practical criterion to identify potential neutrino-emitting blazars. 

\begin{figure}[htbp]
\includegraphics[width=0.5\textwidth]{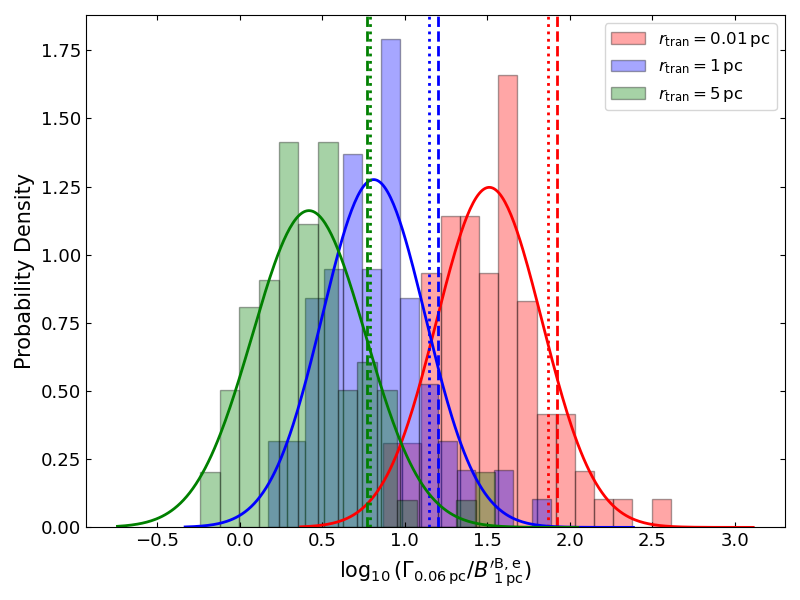}
\caption{Distribution of $\Gamma_{\rm 0.06,pc}/B'^{\rm B,e}_{\rm 1\,pc}$ for the MOJAVE blazar sample. The histograms show the probability density for different transition distances $r{\rm tran}$, indicated by different colors as shown in the legend, while the solid curves denote Gaussian fits to the distributions. The vertical dashed and dotted lines indicate the locations of PKS\,1502+106 and TXS\,0506+056, respectively. Both PKS\,1502+106 and TXS\,0506+056 lie on the high-value side of the distribution across all choices of $r_{\rm tran}$. 
\label{mojave}}
\end{figure}


The results are shown in Fig.\,\ref{mojave} for three representative values of the transition distance, i.e., $r_{\rm tran} = 0.01$, 1, and 5\,pc, respectively. 
The distributions of $\Gamma_{\rm 0.06\,pc}/B'^{\rm B,e}_{\rm 1\,pc}$ are broadly similar under different choices of $r_{\rm tran}$, while both PKS\,1502+106 and TXS\,0506+056 consistently lie on the high-value side of the distributions. Although TXS\,0506+056 is also shown for comparison, it is not included in the quantitative screening because it is not in the MOJAVE sample and the quasi-simultaneous radio observations only provide a lower limit on $\Gamma_{\rm C}$. The values shown for TXS\,0506+056 are therefore obtained using $\Gamma_{\rm C}=30$ as adopted in Table\,\ref{Upara}. The resulting $\Gamma_{\rm 0.06\,pc}/B'^{\rm B,e}_{\rm 1\,pc}$ values are comparable to those of PKS\,1502+106, indicating that adopting PKS\,1502+106 as the reference threshold does not substantially affect the overall screening criterion.
In total, eleven such sources are identified. Excluding PKS\,1502+106, four of them, namely 0133+476, 0202+149, 0605-085, and 1015+359, have clear BLR detections and are classified as FSRQs. We further examine the spatial coincidence of these sources with the IceCat-1 catalog \citep{2023ApJS..269...25A} and find that two sources show positional associations with reported neutrino events. These are 1015+359 (B2 1015+35B), which lies near the edge of the 90\% uncertainty contour of IC-120605A, and 0202+149 (4C\,+15.05), which is spatially associated with IC-150428A. Unfortunately, no X-ray observations were carried out within the one year of the neutrino arrival time. 
Interestingly, our criterion based on $\Gamma_{\rm 0.06\,pc}/B'^{\rm B,e}_{\rm 1\,pc}$ may also naturally explains the lack of neutrino associations with $\gamma$-ray bright FSRQs. Prominent bright FSRQs, such as 3C\,454.3 (2251+158) and CTA\,102 (2230+114), are included in our sample, however their values of $\Gamma_{\rm 0.06\,pc}/B'^{\rm B,e}_{\rm 1\,pc}$ fall around intermediate levels. Consequently, they do not exceed the threshold set by PKS\,1502+106 and are naturally excluded by our proxy threshold. It implies that a high $\gamma$-ray flux alone is insufficient to guarantee efficient neutrino production. 
We also note a very recent observational result, where 0215+015, a source in the MOJAVE sample, has been suggested to be spatially associated with the neutrino event IC-220225A \citep{2026A&A...708A.326E}. Contemporaneous VLBI observations indicate an extreme bulk Lorentz factor as high as $\Gamma \sim 105 \pm 56$. This is much larger than the archival MOJAVE value (see Table\,\ref{tab:mojave_sample}), and is therefore not reflected in our selection shown in Fig.\,\ref{mojave}. Adopting the central value together with the archival magnetic field inferred from MOJAVE, its $\Gamma_{\rm 0.06\,pc}/B'^{\rm B,e}_{\rm 1\,pc}$ would also exceed that of PKS\,1502+106, although the large uncertainty prevents a firm conclusion.
On the other hand, we also specifically examine the extreme beaming sub-sample within the MOJAVE catalog studied in \cite{2025ApJ...991...33P}. Notably, J1733-1304 is spatially associated with the neutrino event IC-160128A and possesses Swift-XRT X-ray observations within one year of the neutrino arrival time. Although it does not exceed the strict $\Gamma_{\rm 0.06\,pc}/B'^{\rm B,e}_{\rm 1\,pc}$ threshold set by PKS\,1502+106, its spatial association and the availability of quasi-simultanesous X-ray data make it a well-motivated target for further study. Consequently, J1733-1304 allow us to apply the same method used for TXS\,0506+056 and PKS\,1502+106 to constrain $Y_{\rm min}$ and the corresponding parameter space of the neutrino-emitting region self-consistently.

\begin{figure}[htbp]
\subfigure{
\includegraphics[width=0.5\textwidth]{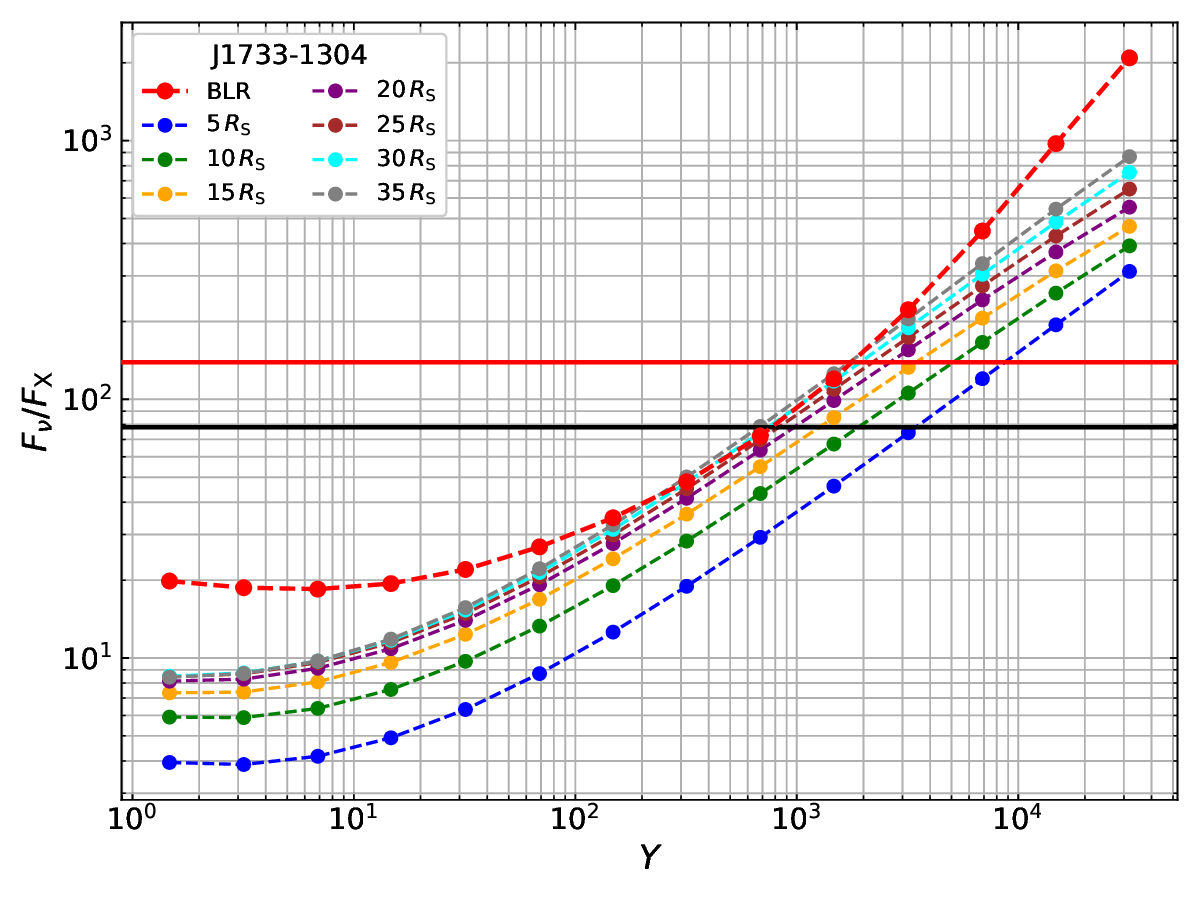}
}
\quad
\subfigure{
\includegraphics[width=0.5\textwidth]{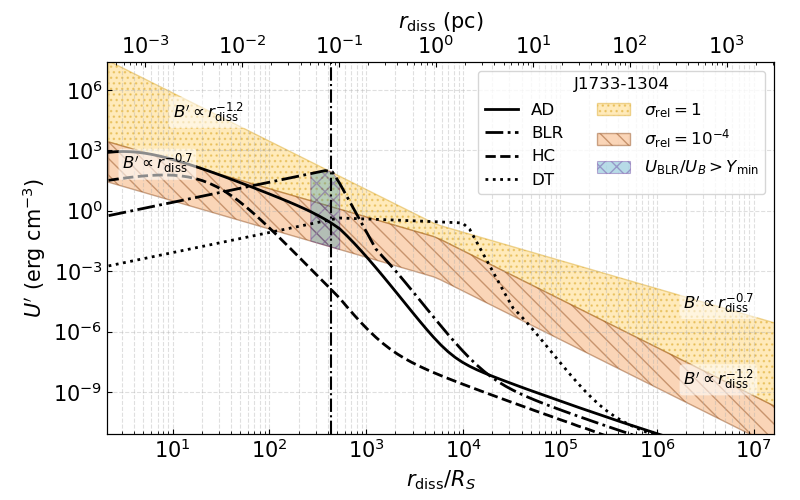}
}
\caption{Upper panel: Ratio of the observed neutrino flux to the integrated X-ray flux as a function of the energy density ratio between the external photon field and the magnetic field for J1733-1304. The line styles have the same meaning as in Fig.\,\ref{Ymin}. Lower panel: Comoving energy densities $U'$ of external radiation fields and the magnetic field as a function of $r_{\mathrm{diss}}$ for J1733-1304. The line styles have the same meaning as in Fig.\,\ref{Urdiss}.
\label{1730}}
\end{figure}

Following the procedure outlined in Sect.\,\ref{ymin}, we first constrain $Y_{\rm min}$ for both the AD-dominated and BLR-dominated scenarios. This requires the neutrino flux distribution and the quasi-simultaneous X-ray photon spectrum. The associated neutrino event IC-160128A is an EHE-gold event with an energy of 583\,TeV. Following the suggestion of \cite{2023ApJS..269...25A}, we adopt a neutrino spectral index of $-2.19$ and assume an integration timescale of one year to derive the expected IceCube neutrino flux distribution. Within one year of the neutrino arrival time, Swift-XRT observations are available, allowing us to construct the corresponding quasi-simultaneous X-ray spectrum after standard data reduction (see Appendix\,\ref{XRT} for details of the XRT analysis). The resulting constraints on $Y_{\rm min}$ are shown in the upper panel of Fig.\,\ref{1730}, yielding $Y_{\rm min}\sim670$--3300. We then apply the condition $Y>Y_{\rm min}$ to locate the potential neutrino production sites. Following Table~\ref{tab:mojave_sample}, we adopt $\Gamma_{\rm C}=31.8$, $r_{\rm C}=63.44\,{\rm pc}$, and $B'^{\rm B,e}_{\rm 1pc}=1.69\pm0.61\,{\rm G}$. To estimate $U'_{\rm ext}(r_{\rm diss})$, we further adopt $L_{\rm AD}=6.8\times10^{45}\,{\rm erg\,s^{-1}}$ and $M_{\rm BH}=2\times10^{9}\,M_{\odot}$ \citep{2016MNRAS.463.3038X}. We find that viable parameter space only exists when $r_{\rm tran}\lesssim0.2\,{\rm pc}$. The lower panel of Fig.\,\ref{1730} presents the case of $r_{\rm tran}=0.1\,{\rm pc}$. It reveals that the condition $Y>Y_{\rm min}$ can only be satisfied in the vicinity of the BLR, with the allowed parameter space remaining rather limited. This result suggests that, unless the jet bulk acceleration is completed on very small scales and the neutrino production region happens to be located within this narrow allowed range, it is highly challenging for J1733$-$1304 to produce detectable neutrino emission efficiently.

Overall, the combination of radio constraints and X-ray observations provides a practical approach to identify promising neutrino-emitting blazars. Radio measurements can be used to construct a proxy for $Y$ and to pre-select candidates, while X-ray data enable a quantitative determination of $Y_{\rm min}$ and the corresponding physical conditions required for neutrino production. This unified framework can be directly applied to large blazar samples and offers a systematic way to connect population studies with detailed modeling of individual sources.

\section{Conclusion} \label{sec:con}
In this work, we investigate the physical conditions required for neutrino production in blazar jets by evaluating the radial profile of the Compton dominance parameter, $Y = U'_{\rm ext}/U'_{\rm B}$, using radio-constrained jet properties. Our results suggest that the neutrino-emitting region is tightly constrained to lie near the BLR, where the external radiation field can be sufficiently enhanced in the jet comoving frame. We show that the critical condition for efficient neutrino production is $Y \gg 1$, which simultaneously ensures a high $p\gamma$ interaction efficiency and suppresses the synchrotron emission from secondary pairs. Such a condition can only be achieved if the dissipation region is located within a limited range of distances, where both the external photon field and the bulk Lorentz factor are sufficiently strong. This places strong constraints on the location, magnetic field strength, and bulk Lorentz factor of the emitting region.

However, the large $Y$ required for neutrino production is generally incompatible with the observed broadband SED if a single emission zone is assumed. This tension naturally implies that the neutrino-emitting region should be physically separated from the dominant electromagnetic emission zone. The specific geometric realizations may vary (e.g., spine--layer or inner--outer models), but the underlying physical requirement remains the same.

An important implication of this picture is that the conditions necessary for neutrino production are realized only in a small subset of blazars. The requirement $Y>Y_{\min}$ can be satisfied either by a small transition distance $r_{\rm tran}$, allowing the jet to reach a sufficiently large bulk Lorentz factor on sub-parsec scales, or by an intrinsically large $\Gamma_{\rm C}$. However, both conditions appear to be uncommon: radio observations suggest that $r_{\rm tran}<R_{\rm BLR}$ is rare, while blazars with extremely large $\Gamma_{\rm C}$ constitute only a small fraction of the population. This naturally explains the rarity of confirmed blazar--neutrino associations. 

While our primary focus has been on the highly beamed blazar population, the methodology developed here is versatile and can be extended to misaligned radio galaxies, provided that high-resolution radio observations are available to constrain the jet kinematics and viewing angles. Future high-resolution radio measurements that can constrain jet structure and dynamics, together with improved neutrino statistics, will be essential for testing this framework and refining the localization of neutrino production sites in blazar jets. In particular, tighter constraints on jet geometry at sub-parsec scales provide a practical avenue for identifying promising neutrino-emitting blazar candidates.

\begin{acknowledgments}
We thank the anonymous referee for insightful comments and constructive suggestions. 
Rui Xue thanks Da-Hai Yan, Ruo-Yu Liu, Jirong Mao, Xiaofeng Li, and Tao An for helpful discussions and comments.
This work is supported by the National Key R\&D Program of China (2023YFB4503300), the National Natural Science Foundation of China (NSFC) under the grants No. 12203043, No. 12203024, and No. 12473020, the Yunnan Province Youth Top Talent Project (Grant No. YNWR-QNBJ-2020-116), and the CAS ``Light of West China" Program. Y.I. is supported by JSPS KAKENHI grant No. JP22K18277 and JP26H00604.
\end{acknowledgments}

\software{NumPy \citep{2020Natur.585..357H}, astropy \citep{2013A&A...558A..33A,2018AJ....156..123A,2022ApJ...935..167A}
          }


\appendix

\section{Secondary pairs spectra with different values of $Y$}\label{sec}
\begin{figure}[htbp]
\subfigure{
\includegraphics[width=0.5\textwidth]{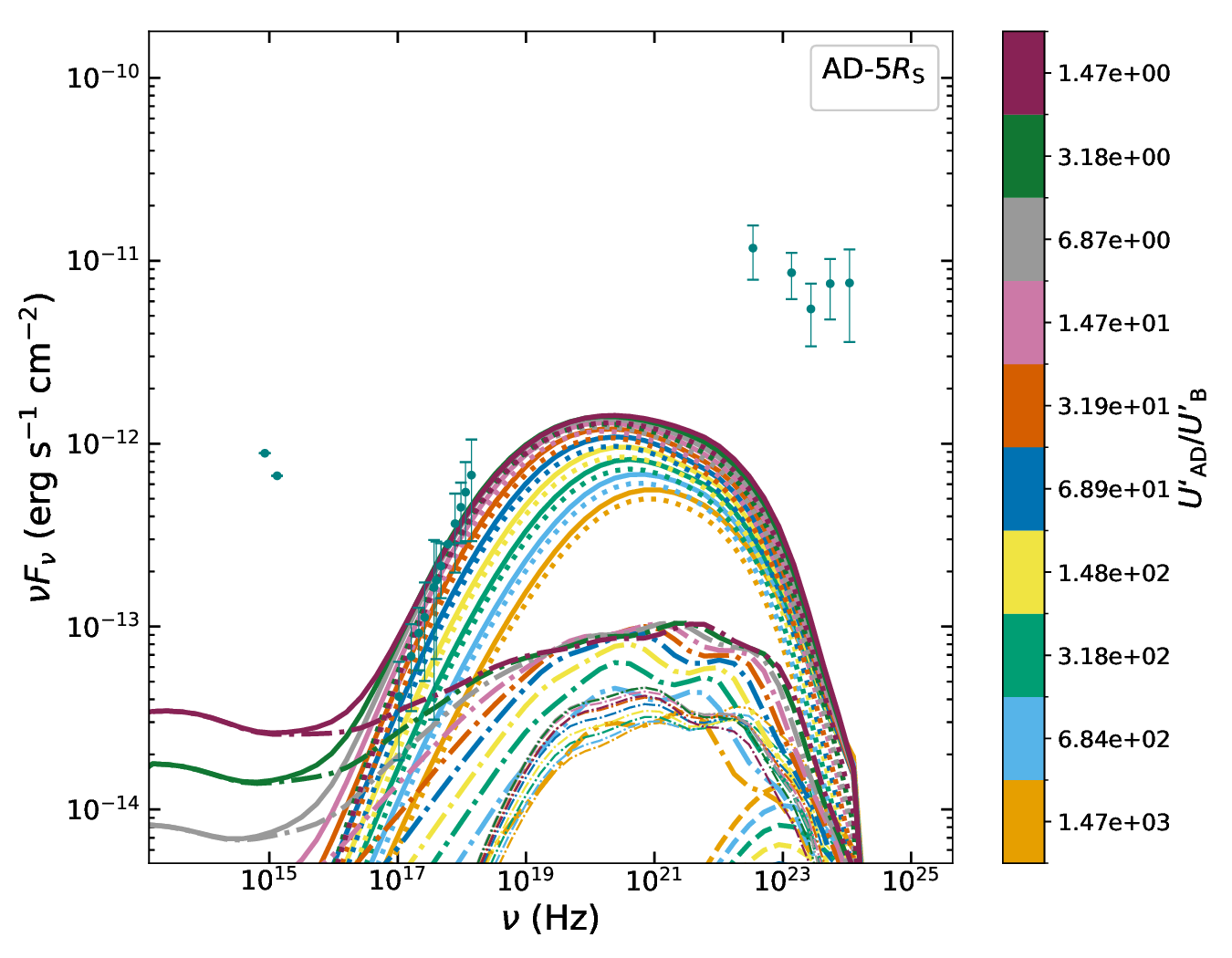}
}
\quad
\subfigure{
\includegraphics[width=0.5\textwidth]{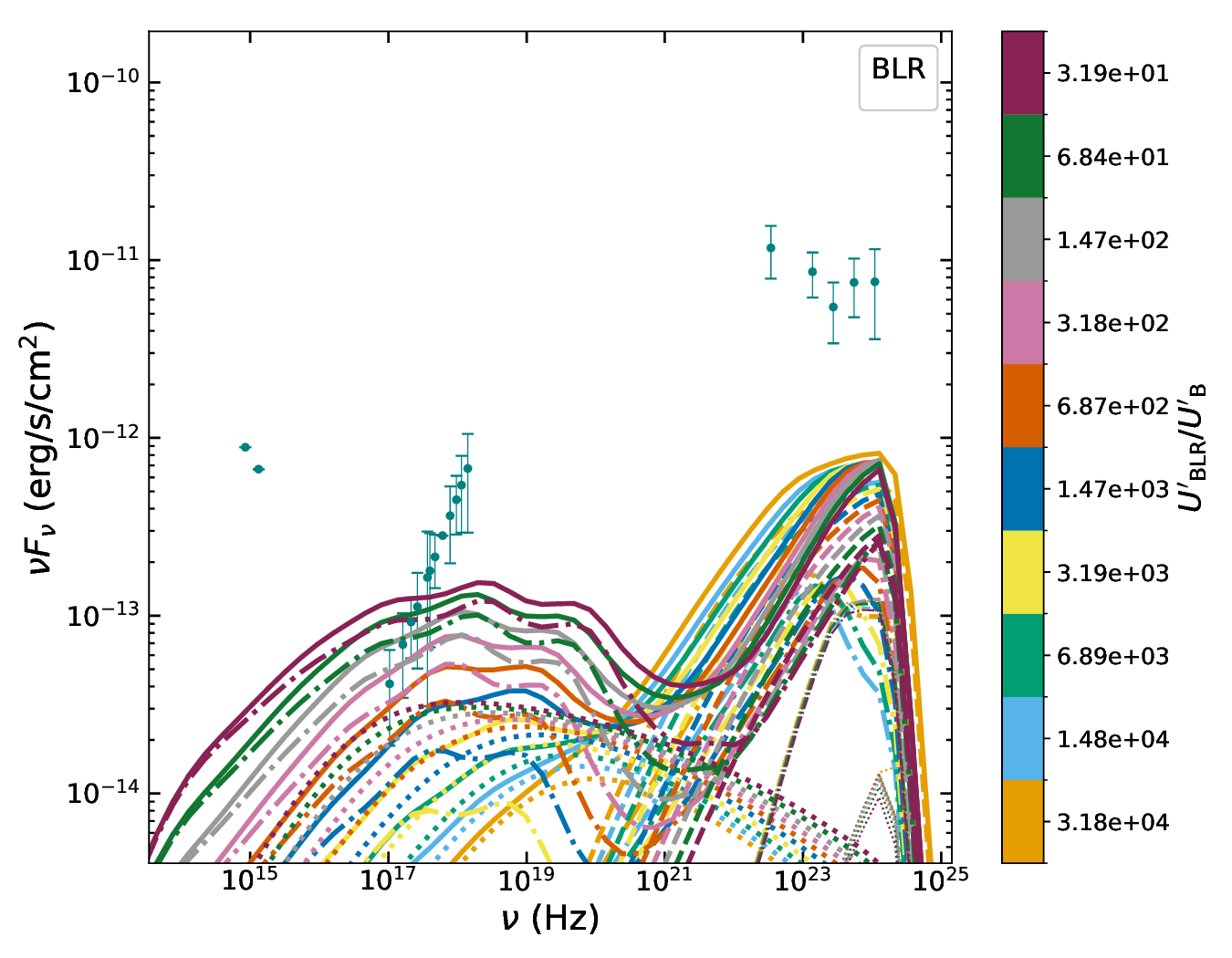}
}
\caption{Secondary pairs spectra of PKS\,1502+106 for different values of $Y$ ($U'/U'_{\rm B}$), shown for the AD-dominated case (left panel) and the BLR-dominated case (right panel). Solid curves represent the total emission, while the dashed, dotted, and dash-dotted curves represent the contributions from pairs produced via $\pi^{\pm}$ decay, the BH process, and internal $\gamma \gamma$ annihilation, respectively. Thick curves correspond to synchrotron radiation, and thin curves represent EC scattering. Different colors represent different values of $Y$, as indicated by the color bar.
\label{app}}
\end{figure}

In Fig.\,\ref{app}, we present representative spectra of the emission from secondary pairs for PKS\,1502+106, corresponding to a dissipation location at $r_{\rm diss}=5\,R_{\rm S}$ (AD-dominated case) and a BLR-dominated case. These examples explicitly illustrate that the spectral shapes of the secondary emission differ significantly between the two scenarios. Specifically, the synchrotron radiation from secondary pairs, which are produced via the BH process (dotted curves) and internal $\gamma\gamma$ absorption (dash-dotted curves), is primarily constrained by the hard X-ray data. In the AD-dominated case, the spectrum of secondary pairs exhibits a slope that closely matches the observed hard X-ray data points, resulting in a relatively low $F_{\nu}/F_{\rm X}$, as reflected by the lower black horizontal line in the bottom panel of Fig.\,\ref{Ymin}. By contrast, in the BLR-dominated case, the low-energy hard X-ray data points already impose a stringent constraint on the secondary emission component. The predicted spectrum does not align with the observed slope, which in turn leads to a higher $F_{\nu}/F_{\rm X}$, corresponding to the higher red horizontal line in the bottom panel of Fig.\,\ref{Ymin}. These differences highlight how the distinct spectral shapes of secondary pairs in the two scenarios translate directly into markedly different observational constraints.

\section{MOJAVE Blazar Sample}
\label{sample}

In Table\,\ref{tab:mojave_sample}, we compile a sample of 82 blazars from the MOJAVE program for which both $\Gamma_{\rm C}$ and $B'^{\rm B,e}_{\rm 1\,pc}$ are available. The radio core distance at 15.4\,GHz, $r_{\rm core,15.4\,GHz}$, is also listed for completeness. This sample is used in Sect.\,\ref{sec:popu} to construct the proxy for $Y$ and to identify potential neutrino-emitting candidates.

\startlongtable
\begin{deluxetable*}{lccc}
\tablecaption{MOJAVE Blazar Sample with Available $\Gamma_{\rm C}$ and $B'^{\rm B,e}_{\rm 1\,pc}$. Sources marked with an asterisk are those with $\Gamma_{\rm 0.06,pc}/B'^{\rm B,e}_{\rm 1,pc}$ exceeding that of PKS\,1502+106. \label{tab:mojave_sample}}
\tablehead{
\colhead{Source} & 
\colhead{$\Gamma_{\rm C}$} & 
\colhead{$r_{\rm core,15.4\,GHz}$ (pc)} & 
\colhead{$B'^{\rm B,e}_{\rm 1\,pc}$ (G)}
}
\startdata
0119+115 & 23.1 & 42.07 & 1.63 \\
0133+476\textsuperscript{*} & 40.1 & 11.6 & 0.77 \\
0149+218 & 15.2 & 35.83 & 1.69 \\
0202+149\textsuperscript{*} & 28.9 & 4.59 & 0.48 \\
0212+735 & 8.1 & 10.65 & 1.27 \\
0215+015 & 37.4 & 37.98 & 1.41 \\
0224+671 & 12.3 & 11.75 & 0.75 \\
0234+285 & 24.7 & 28.67 & 1.71 \\
0333+321 & 18.1 & 34.57 & 1.95 \\
0336-019 & 24.5 & 20.99 & 0.92 \\
0403-132 & 21 & 42.72 & 1.54 \\
0420-014 & 38.2 & 17.37 & 1.41 \\
0528+134 & 21.9 & 28.41 & 1.6 \\
0605-085\textsuperscript{*} & 50.7 & 17.07 & 0.84 \\
0607-157 & 16.5 & 5.6 & 0.68 \\
0716+714 & 24.5 & 6.68 & 0.49 \\
0730+504 & 16.1 & 30.3 & 1.47 \\
0738+313 & 13.9 & 11.85 & 0.81 \\
0748+126 & 18.7 & 16.15 & 0.84 \\
0754+100 & 15.9 & 19.14 & 0.88 \\
0804+499 & 12.1 & 1.51 & 0.48 \\
0805-077 & 42.6 & 114.73 & 2.71 \\
0823+033\textsuperscript{*} & 48.6 & 18.06 & 0.83 \\
0827+243 & 20.6 & 28.16 & 1.18 \\
0829+046 & 11.7 & 4.52 & 0.34 \\
0836+710 & 24 & 42.51 & 1.93 \\
0859-140 & 12.3 & 33.24 & 1.7 \\
0906+015 & 22.1 & 39.67 & 1.61 \\
0917+624 & 14.3 & 17.25 & 1.09 \\
0945+408 & 25.6 & 20.63 & 1.1 \\
1015+359\textsuperscript{*} & 209.7 & 12.65 & 0.92 \\
1036+054 & 6.1 & 7.81 & 0.74 \\
1038+064 & 10.7 & 17.03 & 1.19 \\
1045-188 & 12.8 & 11.19 & 0.86 \\
1101+384\textsuperscript{*} & 1.86 & 0.24 & 0.1 \\
1127-145 & 20.4 & 12.25 & 0.84 \\
1150+812 & 9.3 & 5.75 & 0.68 \\
1156+295 & 24.8 & 32.39 & 1.17 \\
1222+216 & 22.6 & 23.41 & 0.9 \\
1302-102 & 8.6 & 7.28 & 0.7 \\
1308+326 & 27.8 & 24.08 & 0.96 \\
1334-127 & 29.7 & 20.85 & 1.23 \\
1413+135 & 6.4 & 2.1 & 0.44 \\
1458+718 & 8 & 4.38 & 0.51 \\
1502+106\textsuperscript{*} & 34.9 & 8.19 & 0.69 \\
1504-166 & 6.3 & 4.58 & 0.66 \\
1510-089 & 28.8 & 17.71 & 0.73 \\
1532+016 & 7.3 & 17.31 & 1.14 \\
1538+149 & 9.4 & 5.29 & 0.49 \\
1606+106 & 18 & 13.52 & 0.79 \\
1611+343 & 33 & 8.35 & 0.66 \\
1633+382 & 42.4 & 40.67 & 1.62 \\
1637+574 & 11.4 & 9.37 & 0.71 \\
1641+399 & 27.8 & 29.94 & 1.2 \\
1652+398\textsuperscript{*} & 1.63 & 0.22 & 0.1 \\
1655+077 & 16.8 & 7.02 & 0.47 \\
1730-130 & 31.8 & 63.44 & 1.69 \\
1749+096\textsuperscript{*} & 51.4 & 3.11 & 0.33 \\
1749+701 & 5.3 & 10.75 & 1.04 \\
1807+698\textsuperscript{*} & 2.1 & 0.25 & 0.12 \\
1823+568 & 22 & 16.01 & 0.74 \\
1828+487 & 19 & 10.9 & 0.69 \\
1901+319 & 9.2 & 5.39 & 0.87 \\
1928+738 & 8.7 & 6.8 & 0.54 \\
1936-155 & 11.4 & 6.57 & 1.31 \\
2005+403 & 17.3 & 37.61 & 2.34 \\
2113+293 & 7.5 & 4.23 & 1.16 \\
2121+053 & 35.4 & 19.61 & 1.43 \\
2128-123 & 13.9 & 12.05 & 0.98 \\
2131-021 & 21.1 & 19.49 & 1.02 \\
2134+004 & 6.4 & 10.28 & 1.31 \\
2155-152 & 18.7 & 51.02 & 1.89 \\
2200+420\textsuperscript{*} & 11.9 & 0.84 & 0.09 \\
2201+171 & 18.7 & 9.64 & 1.55 \\
2201+315 & 8.7 & 13.96 & 0.95 \\
2223-052 & 27.9 & 27.83 & 1.47 \\
2227-088 & 12.8 & 15.8 & 1.44 \\
2230+114 & 32.7 & 46.7 & 2.12 \\
2243-123 & 9.5 & 7.3 & 0.78 \\
2251+158 & 25.8 & 20.36 & 1.13 \\
2345-167 & 12.2 & 16.18 & 0.9 \\
2351+456 & 62.9 & 30.48 & 1.72 \\
\enddata
\end{deluxetable*}

\clearpage
\section{Data Analysis of Swift-XRT}\label{XRT}
J1733-1304 (also known as NRAO 530) is a highly active X-ray blazar well studied in the literature \citep[e.g.,][]{2006A&A...450...77F}. There are more than 100 visits from the Neil Gehrels Swift Observatory \citep{2004ApJ...611.1005G}. Considering that the corresponding neutrino event was detected on 25th January 2016, we collected the closest data observed by Swift-XRT from April 2016 (for details, see Table \ref{tab:xrt_obs}). The XRT photon counting mode data were analyzed using the FTOOLS software version 6.35. First, the events were cleaned with the {\tt xrtpipeline} task adopting the standard quality cuts. We then extracted the source spectra from a circular region with a radius of 20 pixels, and the background spectra from a larger circle (50 pixels) in a blank area. In the spectral analyses, the ancillary response files from the {\tt CALDB} database were compiled using {\tt xrtmkarf}. Given the limited exposure times of the snapshots, the absorption column density was set to the Galactic value (i.e. $\rm 1.67\times 10^{21}$ $\rm cm^{-2}$, \citealt{2016A&A...594A.116H}), and we performed a joint analysis of the three observations to determine the X-ray photon spectral index. We then carried out an individual analysis for each observation by fixing the photon index derived from the joint analysis. The results of the spectral analyses were also listed in Table \ref{tab:xrt_obs}.

\begin{table*}
\centering
\caption{
Swift-XRT observation log and spectral fitting results for J1733-1304 in 2016.
The columns represent:
(1) the Swift observation ID;
(2) the observation date in UTC;
(3) the effective exposure time;
(4) the net source photon counts;
(5) the unabsorbed X-ray flux in units of ${\rm erg\ cm^{-2}\ s^{-1}}$ between 0.5 and 10~KeV;
(6) the goodness of the fit.
The Galactic hydrogen column density was fixed at
$N_{\rm H}=1.67\times10^{21}\ {\rm cm^{-2}}$,
and the photon index was constrained to
$1.24\pm0.23$
from the joint spectral analysis.
}
\label{tab:xrt_obs}
\begin{tabular}{lccccccc}
\hline
ObsID & Observation Date & Exposure Time & Net Photons & Flux & c-stat/d.o.f. \\
 & (UTC) & (s) &  & ($\log_{10}[{\rm erg\ cm^{-2}\ s^{-1}}]$) & \\
\hline
00035387097 & 2016-04-05 & 921.5  & 53 & $-11.29^{+0.09}_{-0.10}$ & 46/52 \\
00035387098 & 2016-04-12 & 477.0  & 33 & $-11.21^{+0.12}_{-0.14}$ & 38/32 \\
00035387099 & 2016-04-19 & 1099.0 & 49 & $-11.22^{+0.10}_{-0.11}$ & 45/48 \\
\hline
\end{tabular}
\end{table*}


\bibliography{ApJS-rv}{}
\bibliographystyle{aasjournalv7}



\end{document}